\documentclass[twocolumn]{aastex63}

\usepackage{graphicx,subfigure}
\usepackage[utf8]{inputenc}
\usepackage{longtable}
\usepackage{booktabs}
\usepackage{multirow}
\usepackage{savesym}
\savesymbol{tablenum}
\usepackage{siunitx}
\restoresymbol{SIX}{tablenum}
\usepackage{natbib}
\usepackage{amsmath}
\graphicspath{{./}{figures/}}

\font\manual=manfnt at 7pt \def\dbend{\hbox{\raise0.9ex\hbox{\manual\char127\hspace{0.6em}}}}

\providecommand{\e}[1]{\ensuremath{\times 10^{#1}}}

\newcounter{INTERNALionstage}

\def\gtsim{\mathrel{\hbox{\rlap{\hbox{\lower4pt\hbox{$\sim$}}}\hbox{$>$}}}}
\def\lesssim{\mathrel{\hbox{\rlap{\hbox{\lower4pt\hbox{$\sim$}}}\hbox{$<$}}}}

\def\A{{\rm\thinspace \AA}}
\def\cm{{\rm\thinspace cm}}

\def\g{{\rm\thinspace g}}

\def\K{{\rm\thinspace K}}

\def\m{{\rm\thinspace m}}

\def\pcc{{\rm\thinspace cm^{-3}}}

\def\s{{\rm\thinspace s}}

\def\W{{\rm\thinspace W}}

\def\pscm{\mbox{$\cm^{-2}\,$}}

\def\pscm{\mbox{$\cm^{-2}\,$}}

\def\ka{\mbox{{\rm K}$\alpha$}}

\def\h0{\mbox{{\rm H}$^0$}}
\DeclareMathAlphabet{\vib}{OML}{cmm}{m}{it}

\accepted{for publication in ApJ}

\begin{document}

 \title{X-ray spectroscopy in the microcalorimeter era II: A new Diagnostic on Column Density from the Case A to B Transition in H- and He-like Iron
}
\author[0000-0002-4469-2518]{P. Chakraborty}
\affiliation{University of Kentucky \\
Lexington, KY, USA}
\author[0000-0002-8823-0606]{M. Chatzikos}
\affiliation{University of Kentucky \\
Lexington, KY, USA}
\author[0000-0002-2915-3612]{F. Guzm\'an}
\affiliation{University of Kentucky \\
Lexington, KY, USA}
\author{Y. Su}
\affiliation{University of Kentucky \\
Lexington, KY, USA}
\author[0000-0003-4503-6333]{G. J. Ferland}
\affiliation{University of Kentucky \\
Lexington, KY, USA}

\begin{abstract}
 The Soft X-ray Spectrometer (SXS) on board \textit{Hitomi}, with the unprecedented resolving power of R$\sim$1250, allowed the detection of members of the Fe XXV \ka\ complex emission spectra from the center of the Perseus Cluster.  
  In this paper, we introduce a novel method of measuring the column density using the optically thin (Case A) to optically thick (Case B) transition for one- and two-electron systems.  We compare the Fe XXV K$\alpha$ line ratios computed with CLOUDY with that from the \textit{Hitomi} observations 
  in the outer region of the Perseus core using collision strengths from different atomic datasets, and obtain good agreement.
  We also show the effect of turbulence on  Fe XXV K$\alpha$ line ratios and interplay between column density and metallicity.
  Besides, we discuss the atomic number dependence of transition probabilities for allowed and non-allowed transitions, 
  which causes the highly charged He-like systems, such as Fe XXV, to behave fundamentally differently from He I.
  
\end{abstract}

\keywords{galaxies: clusters: individual: Perseus --- X-rays: galaxies: clusters ---
radiative transfer --- galaxies: clusters: intracluster medium}

\section{Introduction}

Galaxy clusters are the largest gravitationally bound objects in the Universe. Due to X-ray emission from the hot ($10^{7}-10^{8}$K) intracluster medium (ICM), they serve as excellent probes of gas dynamics \citep{2008LNP...740....1S, 2009ApJ...703.1297E, 2014PhDT.......227M, 2016mssf.book...93P}, the cooling-heating mechanism \citep{2006PhR...427....1P, 2014Natur.515...85Z} and formation of large scale structure \citep{1995PhR...251....1B, 2006MNRAS.369.1131R, 2012A&ARv..20...54F, 2017AstSR..13...81T, 2018PASP..130f4001C}. Temperature profiles of ICM in many clusters show a temperature drop towards the cluster core, the so-called cool-core clusters \citep{2001ApJ...560..194M}. Radiative cooling time scales in such clusters are significantly shorter than the Hubble time \citep{1992MNRAS.258..177E, 1997MNRAS.292..419W, 2000MNRAS.315..269A}. In principle, this should lead to a slow infall of ICM towards the cluster core, the scenario known as cooling flow \citep{1994ARA&A..32..277F}.
\citet{2003ApJ...590..207P} found a large deficiency in the observed emission from low-temperature X-ray gas, in contrast to the prediction of the standard cooling-flow model.

There are many theories to account for this lack of cool X-ray emission. Feedback by Active Galactic Nucleus (AGN) is the most plausible theory to explain the suppression of cooling in the core \citep{2007ARA&A..45..117M}. Deep Chandra observations provide evidence in favor of AGN feedback through the exchange of mechanical energy between radio-emitting jets and the ICM \citep{doi:10.1146/annurev-astro-081811-125521}. This scenario is backed up by simulations \citep{2010MNRAS.409..985D, 2012ApJ...747...26L}.
Perseus is the brightest X-ray cluster ($z=0.01756$), and the prototypical cool-core cluster. Ariel 5 observations of the Fe XXV and Fe XXVI emission features near 7 keV in Perseus established that its X-ray emission comes from a diffuse hot plasma, permeating the cluster volume \citep{1976MNRAS.175P..29M}. Later, the Fe XXV complex was detected with XMM-Newton at 6.7 keV, and it was used to investigate for the evidence of gas motions in Perseus core \citep{2003ApJ...590..225C, 2004MNRAS.347...29C}. 

Tremendous advancement in X-ray astronomy was made with the launch of the \textit{Hitomi} Observatory on February 17, 2016. The Soft X-ray Spectrometer (SXS) \citep{2016SPIE.9905E..0VK} on board \textit{Hitomi} is equipped with an X-ray micro-calorimeter paired with a Soft X-ray Telescope (SXT). The micro-calorimeter spectrometer provided a resolving power of $R \sim 1250$ \citep{2016Natur.535..117H}, allowing the detection of many weak emission/absorption lines. The previously detected Fe XXV complex in Perseus was resolved by SXS into four main components -- the resonance (w), intercombination (x,y), and forbidden (z) lines (see Table \ref{t:E_Fe XXV}). Also, bulk and turbulent motions of the ICM were precisely measured for the first time with the SXS from Doppler shifts and broadening of spectral lines \citep{2016Natur.535..117H}. 

Future X-ray missions like XRISM (follow-up of \textit{Hitomi}) and ATHENA  will offer a plethora of high spectral resolution X-ray data. Complete computational plasma simulations are required to model several important cluster properties like temperature, turbulence, and metal abundances with high precision. In this paper, we present a study of the Fe XXV X-ray emission in Perseus based on \textit{Hitomi} data with the spectral synthesis code CLOUDY \citep[last reviewed by][]{2017RMxAA..53..385F}. This is a step towards achieving a precise spectral synthesis model for the above future X-ray missions.

With the advancement in spectral resolution, radiative transfer effects that are known in the optical will become accessible to X-ray astronomy. For example, in addition to optically thin line emission propagation (Case A), optically thick emission (Case B) is expected at sufficiently high column densities \citep{1938ApJ....88...52B}. This is the theme of this paper.
Similar effects in the presence of a continuous radiation source \citep[Case C,][]{1938ApJ....88..422B,1999PASP..111.1524F} will be explored in a future paper. 

The Case A/B/C transition is a property of all one- and two-electron systems, although most studies consider H and He in the optical because historically ground-based instruments had the highest throughputs and were able to obtain the highest resolution and S/N spectra. Only now, such works are becoming possible in the X-ray emitting gas from galaxy clusters.
Previous works on X-ray one-electron Case A\&B include \citet{1995MNRAS.272...41S} 
and, for two-electron systems,
\citet{2007ApJ...664..586P}.
In this work, we use CLOUDY to simulate the environment in the outer region of the Perseus core, showing Case A to B transition in the Fe XXV and Fe XXVI emission lines. Varying the column density of the simulated cloud is what drives this transition. Surprisingly, we observe Case A to B transition in both allowed and forbidden lines in the two-electron iron. 

In addition to the Case A to B transfer, another important factor contributing to the change in the emission line intensities
is the loss of identity of a line photon due to line interlocking. As a result of line interlocking, a Fe XXV line photon can be absorbed by Fe$^{23+}$, leading to autoionization in some fraction of the absorbing ion, ultimately destroying the Fe XXV K$\alpha$ photon by Resonance Auger Destruction \citep{2005AIPC..774...99L}. Another atomic process, which we call Electron Scattering Escape (ESE), leads to the change in the Fe XXV K$\alpha$ line intensities at hydrogen column densities greater than 10$^{23}$ cm$^{-2}$. These two processes have been discussed in the first paper of this series, hereafter paper I \citep{2020arXiv200715565C}. However, it is 
difficult to study these processes separately when explaining the variation of the Fe XXV K$\alpha$ line intensities with column density. The prime focus of this paper is on Case A to B transfer, with occasional references to line interlocking and RAD.

The organization of this paper is as follows. Section \ref{Atomic Processes} summarizes all the relevant atomic processes. Section \ref{Case A, Case B} discusses Case A to Case B transition for hydrogen and heavier elements. 
Section  \ref{Line ratios from Hitomi observations} presents our X-ray analysis of the \textit{Hitomi} observational data.
Section \ref{Simulation Parameters} discusses the parameters used for our simulations with CLOUDY. Section \ref{Results} demonstrates our results: optically thin (Case A) to optically thick (Case B) transition in Fe XXV and Fe XXVI, constraints on column density, the interplay between column density and metallicity, and effect of turbulence in Perseus core from the Fe XXV line ratios. Finally, we discuss our results in Section \ref{Discussion}. Unless otherwise mentioned, all uncertainties in this paper are reported at 3 $\sigma$ intervals.

\section{Atomic Processes}\label{Atomic Processes}
In this Section, we describe the various atomic data sources we apply in our spectral modelling. 
Some data sources, especially collisions, are uncertain and we try to estimate this, and show their effects on the spectrum.

\subsection{Energy Levels}
Helium energy levels are taken from \cite{Martin2006}. Energy levels for heavier He-like ions, lithium through zinc, are taken from the CHIANTI atomic database version 5 \citep{2005HiA....13..653L,1997A&AS..125..149D}. 

\begin{figure}[h!]
\centering
\includegraphics[scale=0.32]{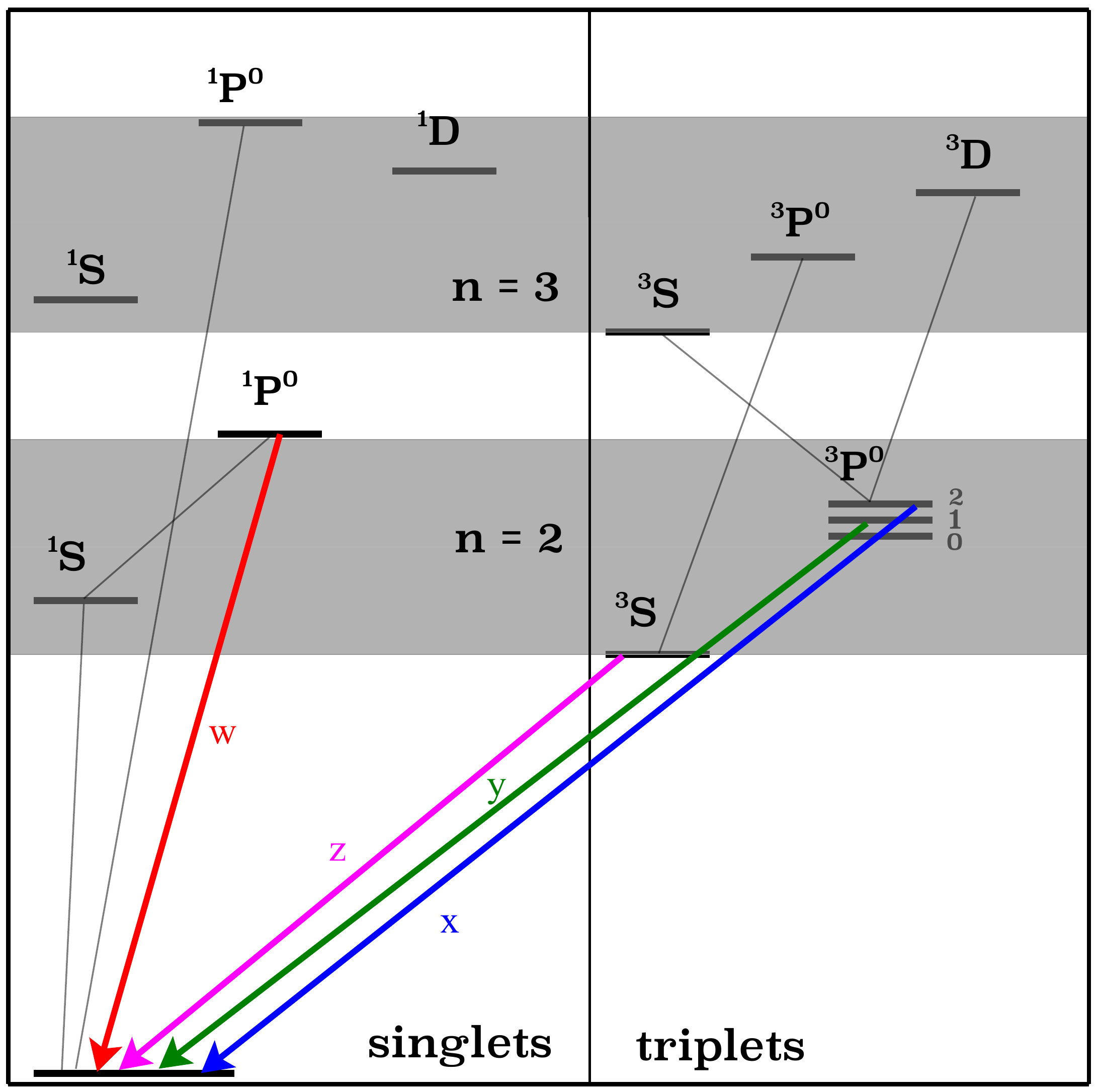}
\caption{Grotrian diagram (not to scale) for He-like ions ($Z\geq 10$). Note that the relative order of the energy levels is different at lower ionic charges. x, y, z, and w transitions are marked with blue, green, purple, and red, respectively.}
\label{f:he_grotrian}
\end{figure}

We use $j$-resolved levels for the $2\,\,^3P$ terms in order to calculate accurate populations for these levels, and  emissivities for the intercombination lines between them and the ground state. We include $nls$ resolved levels for n $\leq$ 5 (where n is the principal quantum number), while levels in the range 5 $\leq$ n $\leq$ 100 are collapsed ( see figure 1 in \citet{2013RMxAA..49..137F} for a schematic representation of resolved and collapsed levels).
Energy levels of He-like ions are represented in Figure \ref{f:he_grotrian}. In Table \ref{t:E_Fe XXV}, there is a list of transition energies from $n=2,3 \to n=1$ for Fe XXV in the interval of 6.5-8.0 keV.

\begin{table*}
\centering{
  \caption{\label{t:E_Fe XXV}List of energies and  transition probabilities for the relevant n=2,3 to n=1,2 transitions in Fe XXV. n=2 levels are $j$-resolved in triplet P.}}
\begin{tabular}{llllrrr}
\hline
\multicolumn{6}{r}{      Transition Probability} \\
\cline{5 -7}
Label & Transition    & Types & Energy (keV) & CLOUDY & NIST & Difference\\
\hline
 w&$2^{1}P\rightarrow1^{1}S$ &E$_{1}$ & 6.7007 & 4.54e+14 &4.57e+14 & 1\%    \\
 x&$2^{3}P_{2}\rightarrow1^{1}S$ &M$_{2}$	& 6.6824  & 6.30e+09 &6.64e+09  &5\%        \\
 y&$2^{3}P_{1}\rightarrow1^{1}S$ & E$_{1}$+M$_{1}$ & 6.6679  & 4.11e+13 &4.42e+13 & 8\%       \\
 z&$2^{3}S\rightarrow1^{1}S$ &M$_{1}$& 6.6368 & 2.15e+08 &2.12e+08&   1\%     \\
 -&$3^{1}P\rightarrow1^{1}S$ &E$_{1}$ & 7.8758 & 1.29e+14 &1.24e+14 & 4\%         \\
 -&$3^{3}P_{1}\rightarrow1^{1}S$ & E$_{1}$+M$_{1}$ & 7.8688 & 1.86e+13 &1.50e+13  &19\%     \\
 -&$3\,\,^{3}P \rightarrow 2\,\,^{3}S$ & E$_{1}$ & 1.2352 & 8.66e+12 & 8.08e+12  & 7\%    \\
 \hline
\end{tabular}
 \end{table*}

\subsection{Radiative transition rates}

\citet{2007ApJ...664..586P} list all the sources of $n=2,3$ to $n=2,1$ transition probabilities used in CLOUDY. 
We show the dependence of transition probabilities for $2\,\,^{1}P\rightarrow1\,\,^{1}S$, $2\,\,^{3}P\rightarrow1\,\,^{1}S$, $2\,\,^{3}S\rightarrow1\,\,^{1}S$, $3\,\,^{1}P\rightarrow1\,\,^{1}S$, and $3\,\,^{3}P\rightarrow1\,\,^{1}S$ transitions on atomic number (Z) in Figure \ref{f:tp}. The significance of this Figure will be discussed in Section \ref{Heavy Elements}.


 Table \ref{t:E_Fe XXV} gives a comparison between transition probabilities from CLOUDY and NIST\footnote{\url{ https://physics.nist.gov/asd } } \citep[version 5.6.1:][]{2018APS..DMPM01004K}. 
They show an agreement within 8\% for $n=2$ to $n=1$ transitions, but a variation up to 19\% for $n=3$ to $n=1$ transitions.


\subsection{Recombination rate coefficients}

Coefficients for radiative recombination (RR) rates are calculated from photoionization cross-sections using the Milne relation \citep{2006agna.book.....O}. 
Photoionization cross-sections for the ground state of He-like ions are taken from \citet{1996ApJ...465..487V}, which uses analytic fits from the Opacity Project. 
Dielectric recombination (DR) data are interpolated from \citet{2006A&A...447..389B}. See \citet{2007ApJ...664..586P}, \citet{2013RMxAA..49..137F}, \citet{2017RMxAA..53..385F} for a discussion on state-specific recombination rates.

\subsection{Collisional data}

\subsubsection{Electron impact $n$-changing collisions}

 We interpolate the collision strengths from various atomic datasets for the right temperature (see Section \ref{Results}) for $n=2-5\rightarrow 1$ transitions of Fe$^{24+}$. 
For higher $n$ and Rydberg levels, we use the {\it ab initio} Born approximation theory by \citet{Lebedev1998}, following the analysis in \citet{Guzman2019}.

\subsubsection{Electron impact bound-free collisions}

We take collisional ionization rate coefficients from \citet{1997ADNDT..65....1V}
for the ground state.  For excited states, rate coefficients are taken from the hydrogenic routines of \citet{1973asqu.book.....A} for the lower temperatures,
and from \citet{1988ApJ...335..516S} at the high-temperature end.



\section{The Case A \& B framework in atoms and highly charged ions}\label{Case A, Case B}
\subsection{Introduction to Case A $\&$ B, H I optical emission}\label{Introduction to Case A, B}
Case A and Case B have been defined for hydrogen emission \citep{1938ApJ....88...52B, 1989agna.book.....O}. Case A refers to an optically thin plasma where Lyman radiation is transmitted without absorption \citep{2006agna.book.....O}. 
At low hydrogen column density, all line photons emitted escape the cloud.

Case B occurs when the ionized cloud becomes optically thick to the H~I Lyman resonance lines, due to high hydrogen column density. 
The Lyman photons undergo multiple scatterings and are degraded to lower energy photons \citep[page 70-71,][]{2006agna.book.....O}. 
For instance, after approximately nine scatterings, a Ly${\beta}$ photon gets degraded to a H${\alpha}$ photon plus two-photon continuum. 
Line formation in most nebulae can be explained by conditions closer to Case B. In this limit, the electron density must be small enough $(n_{e} \leq 10^{8}\pcc)$
for collisional deexcitation rate to be lower than spontaneous emissions \citep{1987MNRAS.224..801H}.


\subsection{Atomic Helium}\label{Atomic Helium}
 Radiative lifetimes for excited helium atoms range between less than a nanosecond to several minutes. 
 For example, $2\,\,^{1}P_{1}$ decays to the ground state in 0.56 ns, which is very short compared to the decay times in $2\,\,^{3}P_{2}$ (3.05 s) and the metastable $2\,\,^{3}S$ levels (131 minutes).
These are respectively examples of E$_{1}$ (electric dipole), 
M$_{2}$ (magnetic quadrupole), and M$_{1}$ (magnetic dipole)  transitions
for $n= 2\rightarrow 1$. The decay of $2\,\,^{3}P_{1}$ to the ground state has a radiative lifetime of $\sim$ 5.68 ms and is a combination of E$_{1}$ and M$_{1}$ \citep{kunze2009introduction}.

In Figure \ref{f:tp} we plot transition probabilities 
for He-like ions with $2 \leq Z \leq 30$. 
From the Figure it is clear that in helium, there is no fast transition to the ground from excited triplet states due to their longer radiative lifetimes (and smaller transition probabilities ($A_{u,l}$)). Optical thickness is proportional to the absorption cross-section at frequency $\nu$ ($\alpha_\nu$), which is proportional to $A_{u,l}$ (refer to equation \ref{e:2} and \ref{eqn:LineAbsorptionCrossSection}). Therefore, for helium, there will be no Case A (optically thin) to Case B (optically thick) transition for triplet to ground transitions.
Fast transitions only occur in the allowed singlet to singlet transitions
(E$_{1}$) with decay times of the order of nanoseconds and large absorption
cross-sections, thus allowing for Case A to B transitions. 
This is not the case, however, for higher atomic numbers. Atomic number dependence of transition probability for allowed and forbidden transitions will be discussed in Section \ref{Heavy Elements}.

\subsection{X-ray emission from He-like Fe XXV}\label{Heavy Elements}
Along the two-electron iso-sequence, the next abundant elements are
C, O, and N, emitted in soft X-rays and discussed by 
\citet{2007ApJ...664..586P}. This paper continues to iron, inspired by recent \textit{Hitomi} observations of the Fe K$\alpha$ complex in the Perseus galaxy cluster.

In Fe XXV, in contrast to the atomic helium discussed in Section
\ref{Atomic Helium}, triplet to singlet transitions can also become 
optically thick (Case B) along with the singlet to singlet transitions. 
This is surprising at first sight, but can be explained with the 
transition probability ($A_{ki}$) dependence on the atomic number (Z) 
for different types of transitions.   

Probabilities for E$_{1}$ transitions ($2\,\,^{1}P-1\,\,^{1}S$ (w),     
 $3\,\,^{1}P-1\,\,^{1}S$) grow with the fourth power of Z 
 \citealp[(A$_{ki}$ $\propto Z^{4}$,][]{2002ApJS..141..543J}), whereas transition probabilities for M$_{1}$ ($2\,\,^{3}S-1\,\,^{1}S$ (z)) and M$_{2}$ ($2\,\,^{3}P_{2}-1\,\,^{1}S$ (x)) grow much faster with Z \citep[(approximately $\propto Z^{10}$  \rm {and}  $  \propto Z^{8}$,][]{1977PhRvA..15..154L}). The transition probabilities for $2\,\,^{3}P_{1}-1\,\,^{1}S$ (y) and $3\,\,^{3}P_{1}-1\,\,^{1}S$ are a combination of E1 and M1, and also grow much faster than the allowed E$_{1}$ transitions.
Therefore in higher Z ions like iron,  transition probabilities ($A_{u,l}$) and  optical depths  for  some of the triplet to singlet transitions become comparable to that of the allowed singlet to singlet transitions (see Figure \ref{f:tp}). This implies that Case A to B transition occurs in singlet to singlet as well as triplet to triplet transitions in Fe XXV and other higher Z He-like ions.

\subsection{ X-ray emission from H-like Fe XXVI}\label{X-ray emission from H-like Fe XXVI}
Unlike helium, there is no fundamental difference between atomic hydrogen 
and the highly charged H-like ions. 
This point is elaborated in Section \ref{Fe_XXVI}.

\begin{figure}[h!]
\centering
\includegraphics[scale=0.25]{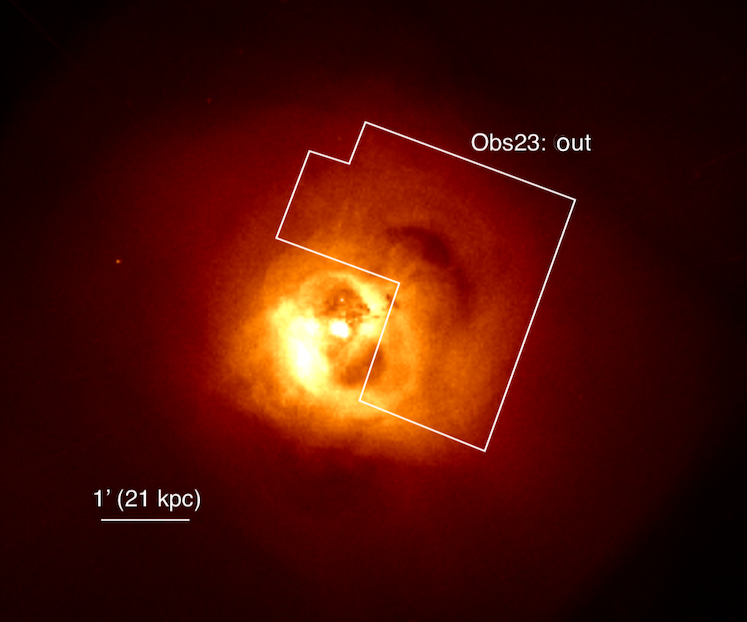}
\caption{The outer region of Perseus core from Hitomi SXS observation (marked with Obs23 : out) overlaid on a Chandra X-ray image of Perseus. 
}
\label{fig:Perseus}
\end{figure}

\begin{figure}
\subfigure{\includegraphics[width = 3in]{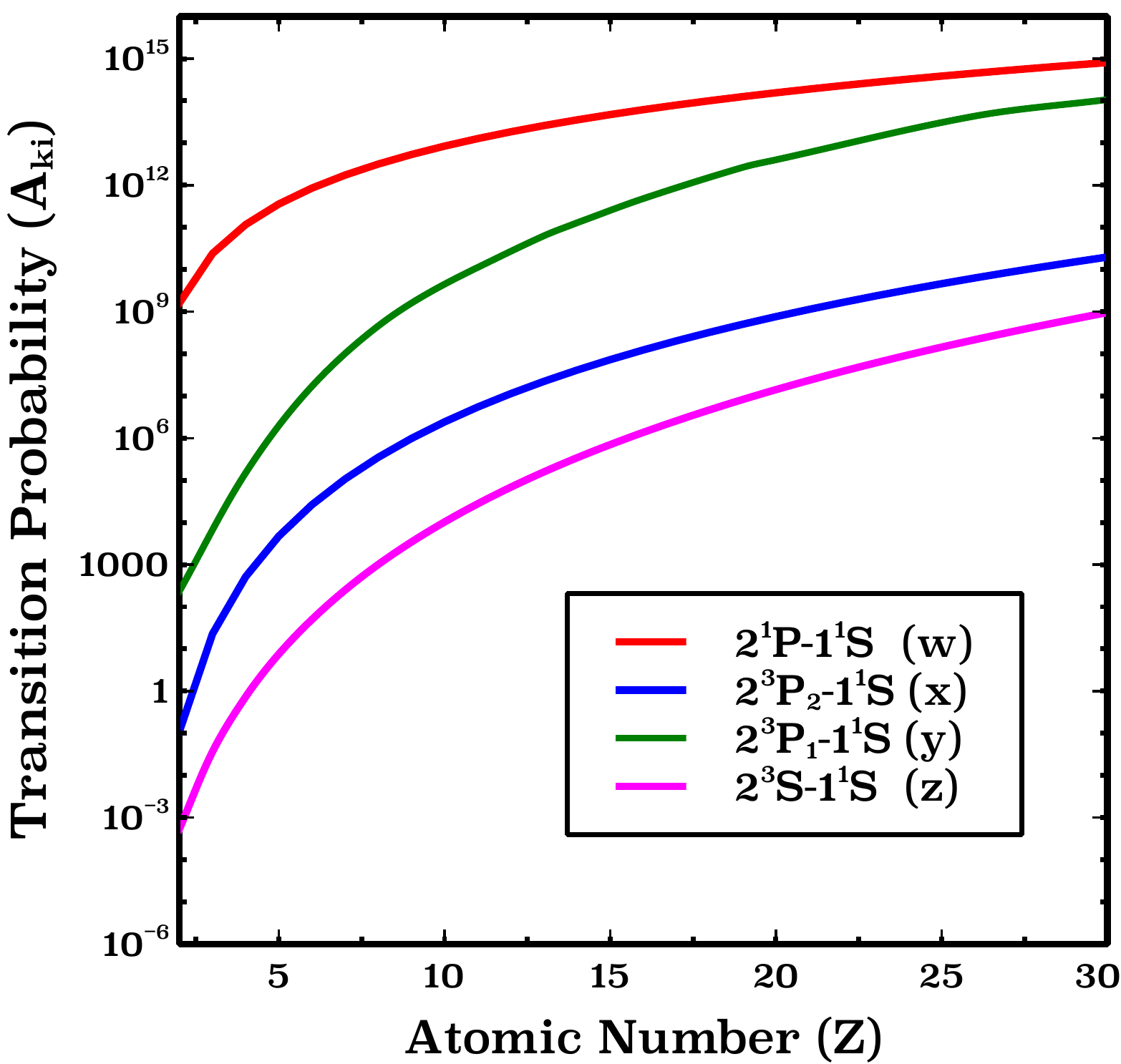}}\hspace{2cm}
\subfigure{\includegraphics[width = 3in]{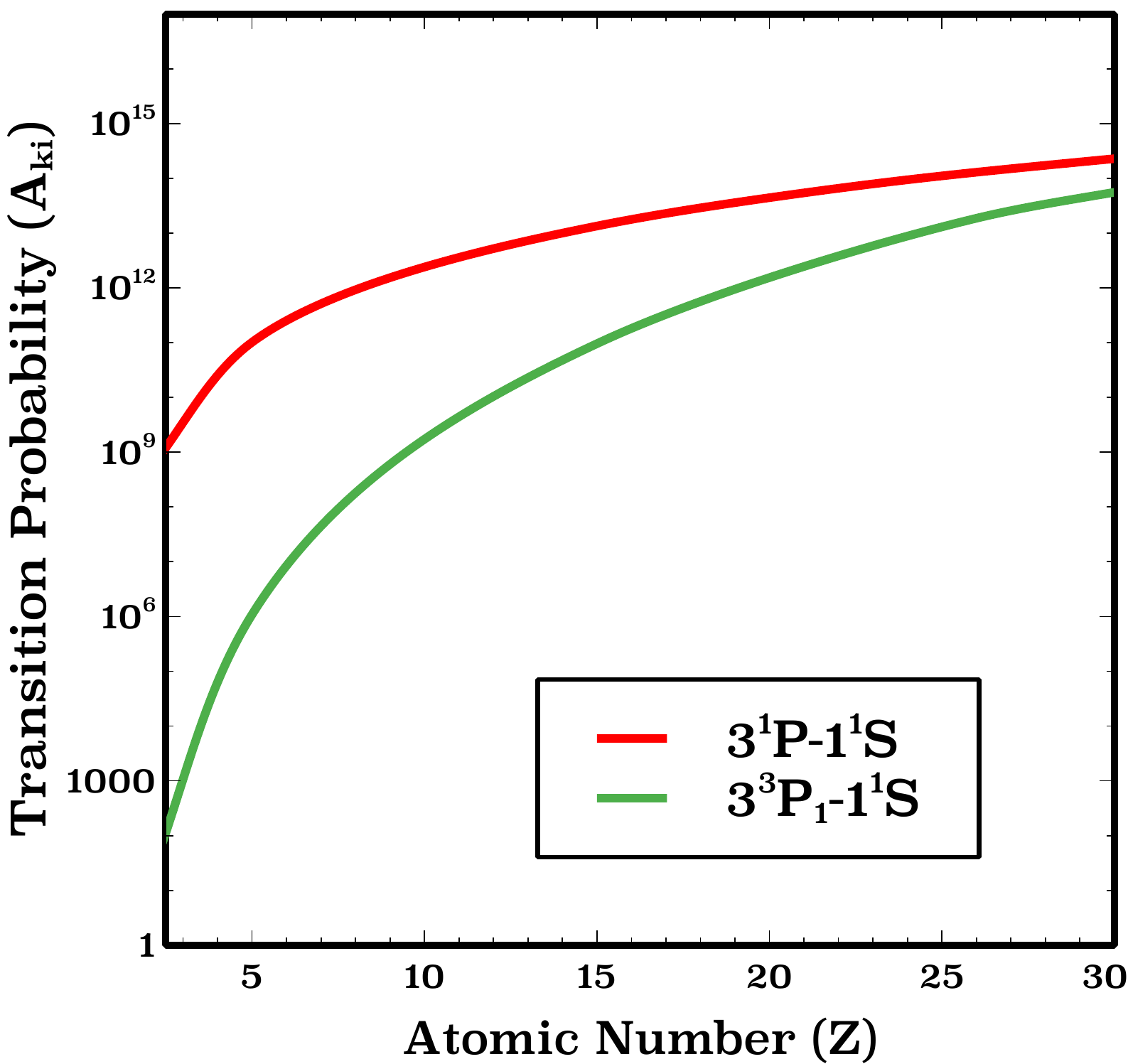}} 
\caption{Top: Transition probabilities vs. Z for certain n=2 to n=1 transitions for He-like ions.
      Bottom: Transition probabilities vs. Z for certain n=3 to n=1 transitions.}
\label{f:tp}
\end{figure}

\section{Line ratios from \textit{Hitomi} observations}\label{Line ratios from Hitomi observations}
\begin{figure}[h!]
\centering
\includegraphics[scale=0.35]{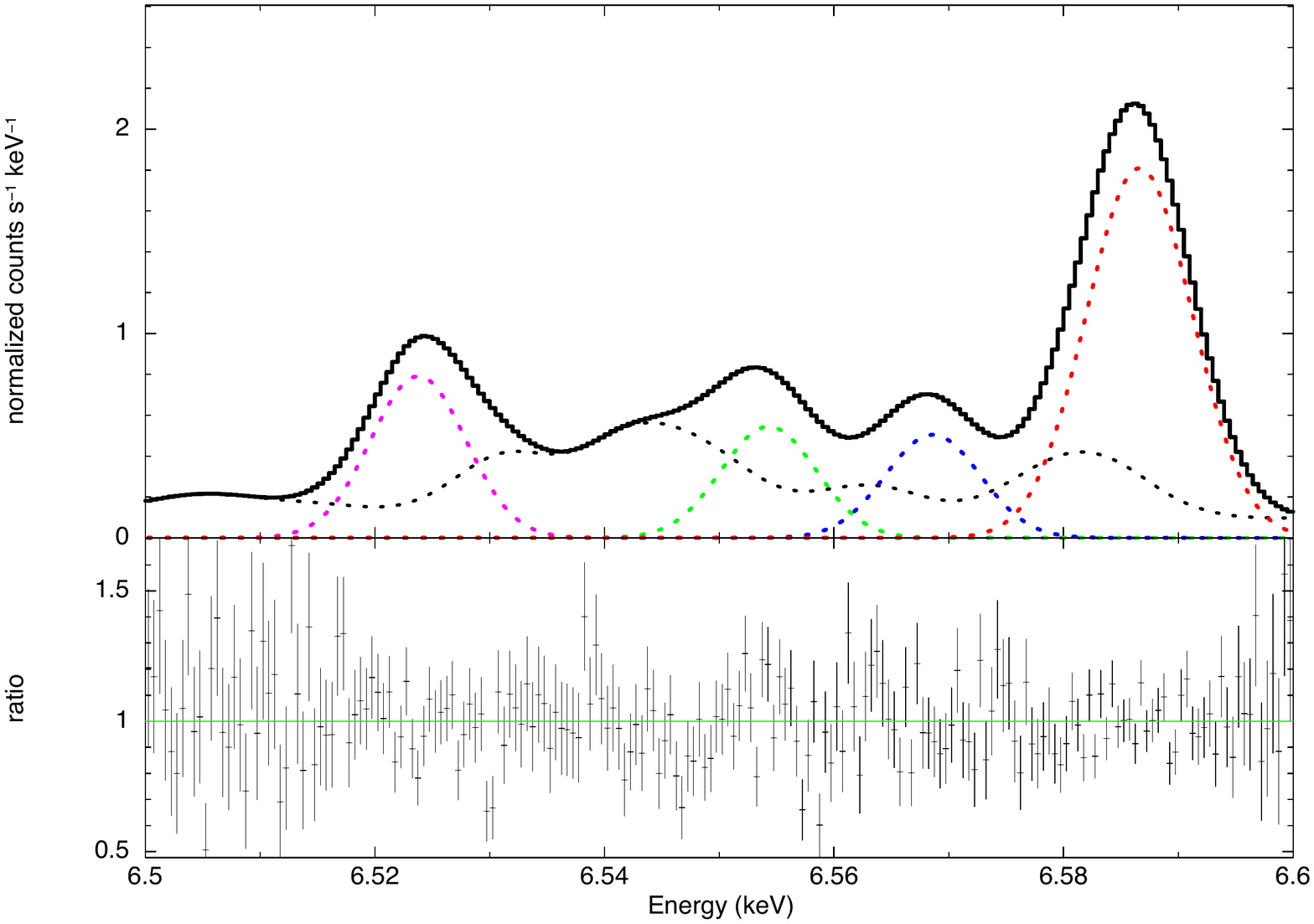}
\caption{The black solid line shows the best-fitting model with x, y, z, w emissivities set to zero, and four added Gaussians centered at their redshifted laboratory energies. Red, blue, green, purple dotted lines show the four Gaussians for w, x, y, and z separately. The black dotted line shows the {\tt bvvapec} model without x, y, z, and w.
}
\label{fig:xspec}
\end{figure}

The observational spectra for the Fe XXV K$\alpha$ is extracted using HEAsoft
version 6.25\footnote{\url{https://heasarc.gsfc.nasa.gov/docs/software/heasoft/}} in the outer region of Perseus core 
following the \textit{Hitomi} step by step analysis guide\footnote{\url{https://heasarc.gsfc.nasa.gov/docs/hitomi/analysis/hitomi_analysis_guide_20160624.pdf}}.
We use the following observations with observation IDs:
100040020, 100040030, 100040040, and 100040050 for extracting the spectra for the region marked with Obs23:out in \citet{2018PASJ...70...10H} (we show this region in Figure \ref{fig:Perseus}).
Event files for these observations are combined and filtered with the Xselect package. Four NXB spectra are extracted separately for the four event files with \texttt{sxsnxbgen}, and averaged using \texttt{mathpha}. RMF, exposure map, and ARF were generated using \texttt{sxsmkrmf}, \texttt{ahexpmap}, and \texttt{aharfgen} respectively.
For fitting the spectra, we use Xspec version 12.10.1 \citep{1996ASPC..101...17A}.

The X-ray emission from our region of interest was modeled as a velocity broadened single-temperature collisionally ionized plasma with variable 
element abundances ({\tt bvvapec}), attenuated by the cold matter absorption 
in our Galaxy ({\tt TBabs}). 
The absorbing hydrogen column density was set to 1.38$\times$10$^{21}$ cm$^{-2}$ \citep[Leiden/Argentine/Bonn Survey of Galactic HI:][]{2005A&A...440..775K}.
The {\tt bvvapec} model was modified whenever necessary by setting the emissivities
to zero for selected lines, following the addition of corresponding Gaussian components. 
The contamination by the central AGN emission is negligible in the outer region;
thus such effects were not included in our model. 
 
 We use the outer region temperature (4.05 $\pm$ 0.01 keV),
 Fe abundance (0.65 $\pm$ 0.01), and turbulent velocity (141 $\pm$ 5 km/s)
 from a broad-band fit in the energy range 1.8-20.0 keV from
 \citet{2018PASJ...70...10H} (errors are reported at 1$\sigma$ confidence level). Solar abundance table by \citet{2009M&PSA..72.5154L} 
 was used throughout. For calculating the line fluxes for x, y, z, and w,
we set the line emissivities for these lines to zero, and add four Gaussians with energies centered at their
redshifted laboratory energies
(E$_{0x}$= 6.68245 keV , E$_{0y}$= 6.66790 keV , E$_{0z}$= 6.63684 keV , E$_{0w}$= 6.70076 keV). 
In the modified {\tt bvvapec} model: {\tt tbabs}*({\tt bvvapec} + {\tt zgauss$_{x}$} + {\tt zgauss$_{y}$} + {\tt zgauss$_{z}$} + {\tt zgauss$_{w}$}), 
we tie the line widths of x, y, z together, 
and set the line width for w free to vary. 
This is because w  is reported to be slightly broader than 
the other three lines in Fe XXV K$\alpha$ complex \citep{2018PASJ...70...10H}.
Normalization of the four Gaussians were set free, and redshifts were tied together.
 Our best-fit model for the observed spectra for Fe XXV K$\alpha$ complex from Obs23:out is  shown in Figure \ref{fig:xspec} with a black solid line. Red, blue, green, and purple dotted lines show the four Gaussians for w, x, y, and z, respectively. The black dotted line shows the {\tt bvvapec} model with x, y, z, and w emissivities set to zero. In such conditions, major contributions to the {\tt bvvapec} model come from Fe XXIV satellite lines, following the contributions from Cr XXIII, and Fe XXIII in the energy range of the Figure. 
 Best-fit redshift for our model is $z = 0.0173308^{+2.53e-05}_{-2.46e-05}$. 
 Best-fit parameters for the four Gaussians and line fluxes 
 with the line ratios are listed in Tables \ref{t:linefluxbestfit} and \ref{t:lineflux}, respectively.

\begin{table*}
\centering{
  \caption{\label{t:linefluxbestfit}Best-fit normalization and sigma of the 
  four Gaussians from our model: {\tt tbabs}*({\tt bvvapec} + {\tt zgauss$_{x}$} + {\tt zgauss$_{y}$} + {\tt zgauss$_{z}$} + {\tt zgauss$_{w}$}).
  Temperature, Fe abundance, and turbulent velocity in the {\tt bvvapec} 
  model are set to the values from broad-band fits in Obs23:out. 
  Line emissivities of x, y, z, and w are set to zero.}}
\begin{tabular}{cccc}
\hline
Label & Normalization ($\times 10^{-5}$ photons/cm$^{-2}$/s) & Sigma (keV)\\
\hline
x & 9.80$^{+0.92}_{-0.88}$ & 3.61e-03$^{+2.78e-04}_{-2.63e-04}$ \\
\\
y & 10.56$^{+1.03}_{-0.98}$ & 3.61e-03$^{+2.78e-04}_{-2.63e-04}$\\
\\
z & 15.34$^{+1.10}_{-1.07}$& 3.61e-03$^{+2.78e-04}_{-2.63e-04}$\\
\\
w & 39.22$^{+1.56}_{-1.52}$ & 4.17e-03$^{+1.94e-04}_{-1.87e-04}$\\

\hline
\end{tabular}
 \end{table*}

\begin{table*}
\centering{
  \caption{\label{t:lineflux}List of net line fluxes and ratios with w
  for the lines in Fe XXV K$\alpha$ complex from the best-fit model. 
  Line fluxes and their errors are reported in units of erg cm$^{-2}$ s$^{-1}$.}}
\begin{tabular}{ccccccc}
\hline
Label & Line flux  & Flux error(positive)&Flux error (negative) &Ratio with w & positive error&negative error \\
\hline
 x & 1.01e-12 & 9.46e-14 & 9.14e-14 & 0.249&  2.50e-02 & 2.47e-02 \\
 y & 1.09e-12 & 1.08e-13 & 9.98e-14 & 0.268 & 2.91e-02&  2.63e-02\\
 z & 1.57e-12 & 1.13e-13 & 1.09e-13 & 0.387 & 3.13e-02 & 3.11e-02\\
 w & 4.06e-12 & 1.61e-13 & 1.58e-13&-&-&-\\
 
\hline
\end{tabular}
 \end{table*}

\section{Simulation Parameters}\label{Simulation Parameters}

For the CLOUDY calculations, we simulate the environment 
in the outer region of Perseus core, introduced as Obs23:out in 
Section \ref{Line ratios from Hitomi observations}. We choose the temperature inside the error interval of the observation: 4.05$^{+0.01}_{-0.01}$ keV. The hydrogen density is set at 0.03 cm$^{-3}$.
Note that, the hydrogen density drops with increasing distance from the cluster core.
For simplicity, we assume an average hydrogen 
density of 0.03 cm$^{-3}$ for our region of interest (30-60 kpc). 
The radial dependence of hydrogen density will be explored in future papers. 

We set the Fe abundance to be in the range of  0.65$^{+0.01}_{-0.01}$ of solar, as discussed
in Section \ref{Line ratios from Hitomi observations}.
Sections \ref{Interplay between column density and metallicity},
where we show the variation of line ratios with changing metallicity, is an exception.
We use the solar abundance table provided by \citet{2009M&PSA..72.5154L} to match with \citet{2018PASJ...70...10H}. 
The turbulence is set to 150 km/s (this choice is elaborated
later in Section \ref{Effects of turbulence} ) for all our calculations 
except for Section \ref{Effects of turbulence}, where we use two additional values for turbulence.
In Section  \ref{Effects of turbulence} and partly in Section \ref{Interplay between column density and metallicity}, hydrogen column density is fixed  at the reported value by \citet{2018PASJ...70...12H} ($ N_{\rm H,hot} \sim  1.88\times 10^{21} \pscm$). 
This is the column density of hot absorbing gas, derived from a best-fit baseline model defined with SPEX \citep{1996uxsa.conf..411K}. In their model $N_{\rm H,hot}$ is set as a free parameter, along with temperature and turbulent velocity of hot gas, emission measure, redshift, and abundances of Si, S, Ar, Ca, Cr, Mn, Fe, and Ni.

For the collision strengths, we use the recent calculations 
of \citet{2017A&A...600A..85S}, who adopted the independent process 
and isolated resonances approximation using distorted waves (IPIRDW).
Since we find instances of disagreement between observed and calculated line ratios, 
we also check the effect of different atomic datasets on the calculated
line ratios (such as Figure \ref{f:col_den} and Figure \ref{fig:Metallicity}).
These include collision data from \citet{2001JPhB...34.3179W},
A. D. Whiteford (2005) \footnote{ADF04, OPEN-ADAS database, helike\_adw05\#fe24.dat, Website: \url{https://open.adas.ac.uk/}},  and A. Giunta (2012) \footnote{ADF04, ls\#fe24.dat, OPEN-ADAS database}. 
 \citet{2018PASJ...70...10H} shows the presence of resonance scattering (RS) in Perseus core, which results in flux suppression in w. 
In the outer region (Obs23:out), the flux suppression factor in w is reported to be $\sim 1.28$. 
We make a correction for this factor for our calculations, as we use a simple model with plane-parallel geometry
that does not account for the photons
lost due to resonance scattering in our line of sight. This will be further discussed in Section \ref{Discussion}.

\section{Results}\label{Results}
\subsection{Case A \& Case B in Iron}\label{Case A,Case B in Iron}
 
 \subsubsection{Fe XXV}\label{Fe XXV}

The Case A to B transition happens when the total line-center optical depth of the Lyman lines becomes $\ge 1$. 
Note that the variation in optical depth is determined by the column density of that species, Fe XXV, in this case. 
The column density of Fe$^{24+}$ (N(Fe$^{24+}$)) can be calculated from 
the hydrogen column density (N(H)) for a given metal abundance (Fe/H), and ionization fraction (Fe$^{24+}$/ Fe) with the following equation:
\begin{equation}\label{e:1}
\textrm{N(Fe$^{24+}$)} =  \left(\frac{\textrm{Fe$^{24+}$}}{\textrm{Fe}}\right)\left( \frac{\textrm{Fe}}{\textrm{H}}\right)   \textrm{N(H)}
\end{equation}
This conversion is shown in the top x-axes in Figure  \ref{f:CaseAB2}, \ref{f:3to1}, and \ref{fig:3to1opt}.

\begin{figure}[h!]
\centering
\includegraphics[scale=0.46]{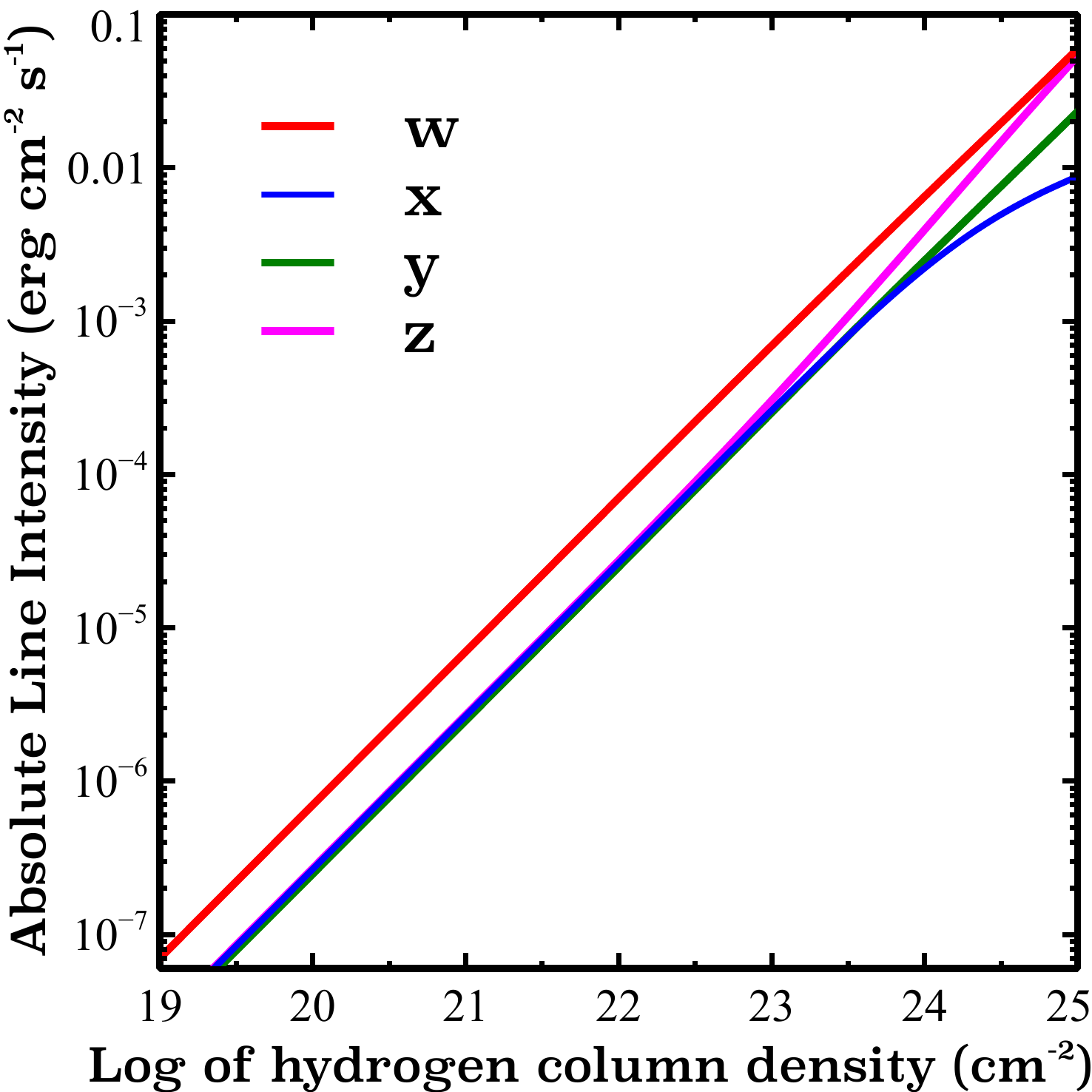}
\caption{Absolute line intensities of  x, y, z, and w vs. log of hydrogen column density. }
\label{f:CaseAB1}
\end{figure}


Decay rates in Fe$^{24+}$ for triplet to singlet  and singlet to singlet  transitions are comparable to each other for some transitions, 
 as discussed in Section \ref{Heavy Elements}. 
 This allows the transfer from the optically thin (Case A) to the optically thick regimes (Case B) for singlet to singlet,
 as well as triplet to singlet  transitions. 

 
Figure \ref{f:CaseAB1}, taken from paper I, shows the line intensities for x, y, z, and w in the Fe XXV K$\alpha$ complex. 
The continuous increase in the line intensities with  column density makes it difficult to detect 
where the Case A to B transition occurs for the individual spectral lines.
To demonstrate the Case A to B transition,
 it is best to consider the variation of line ratios with column density.
Among the four members in Fe XXV K$\alpha$ complex, w is the first line to become optically thick, followed by y, x, and z (Refer to Figure 1b in paper I).  z only becomes optically thick at very high column densities (N$_{H}$ $\geq$ 10$^{25}$ cm$^{-2}$). The Figure also gives an estimate of what fraction of the optical depth  for the four lines comes from their single-line optical depth solely (in this case absorption by Fe$^{24+}$ only ). The total line-center optical depths in x, y, z, and w respectively have $\leq$ 1\%, $\sim$ 60\%, $\leq$ 1\%, and $\sim$ 100\% contribution from absorption by Fe$^{24+}$. 
As z is the last line to become optically thick, we plot line ratios of x, y, and w  relative to z with 
 hydrogen column density (see Figure \ref{f:CaseAB2}).


In the lower column densities, the lines ratios are parallel to each other
and remain constant with increasing column density. 
This is because the individual line intensities are all parallelly
increasing when the column density, and therefore optical depth, is small (Case A). 

At higher column densities (Case B limit), the Figure shows an overall decrease in w/z, x/z, and y/z, which can be caused by the decrease in the numerator, or increase in the denominator, or a combination of these two. In between these two cases, there is a transition region representing
the transfer from Case A to Case B.

The probability that an x photon gets absorbed by Fe$^{24+}$ itself is $\leq$ 1\%.  The  effect of Case A to B transfer in x following absorption by Fe$^{24+}$ is, therefore, very minimal. Apart from this, contributions from RAD and ESE cause a deficit in the x line intensity (paper I).



 \begin{figure}[h!]
\centering
\includegraphics[scale=0.48]{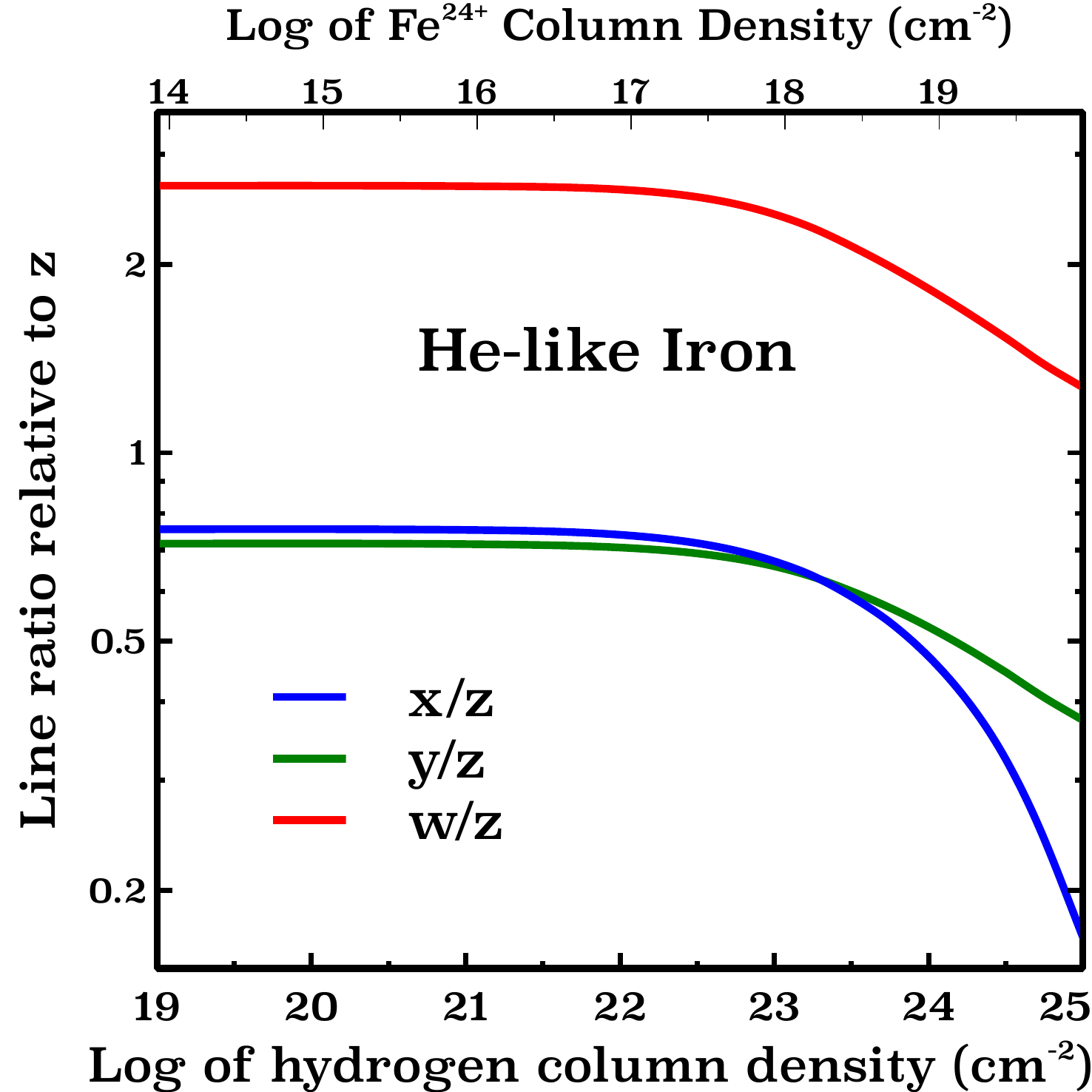}
\caption{Fe XXV  line  ratios  with respect to z for $2\rightarrow 1$ transitions vs log of hydrogen column density.
 The x-axis on top shows log of Fe$^{24+}$ column density. }
\label{f:CaseAB2}
\end{figure}

\begin{figure}[h!]
\centering
\includegraphics[scale=0.4]{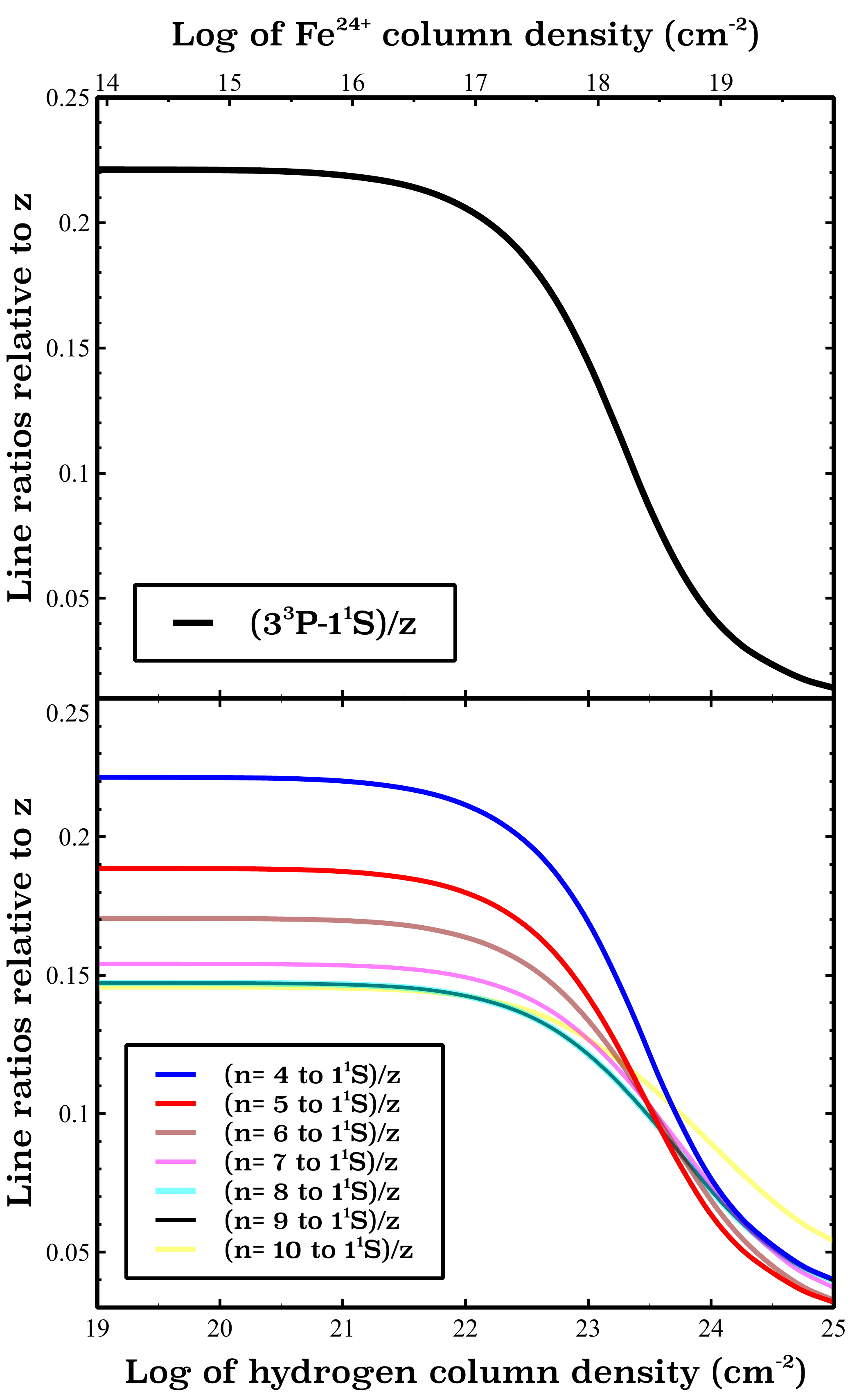}
\caption{Upper panel: Fe XXV  line  ratios with respect to z for $3\,\,^{3}P \rightarrow 1\,\,^{1}S$  transition vs. log of hydrogen column density. Lower panel: Fe XXV  line  ratios with respect to z for $n= 4, 5...\rightarrow 1\,\,^{1}S$ transitions vs. log of hydrogen column density.
The x-axis on top shows log of Fe$^{24+}$ column density.}
\label{f:3to1}
\end{figure}

\begin{figure}[h!]
\centering
\includegraphics[scale=0.5]{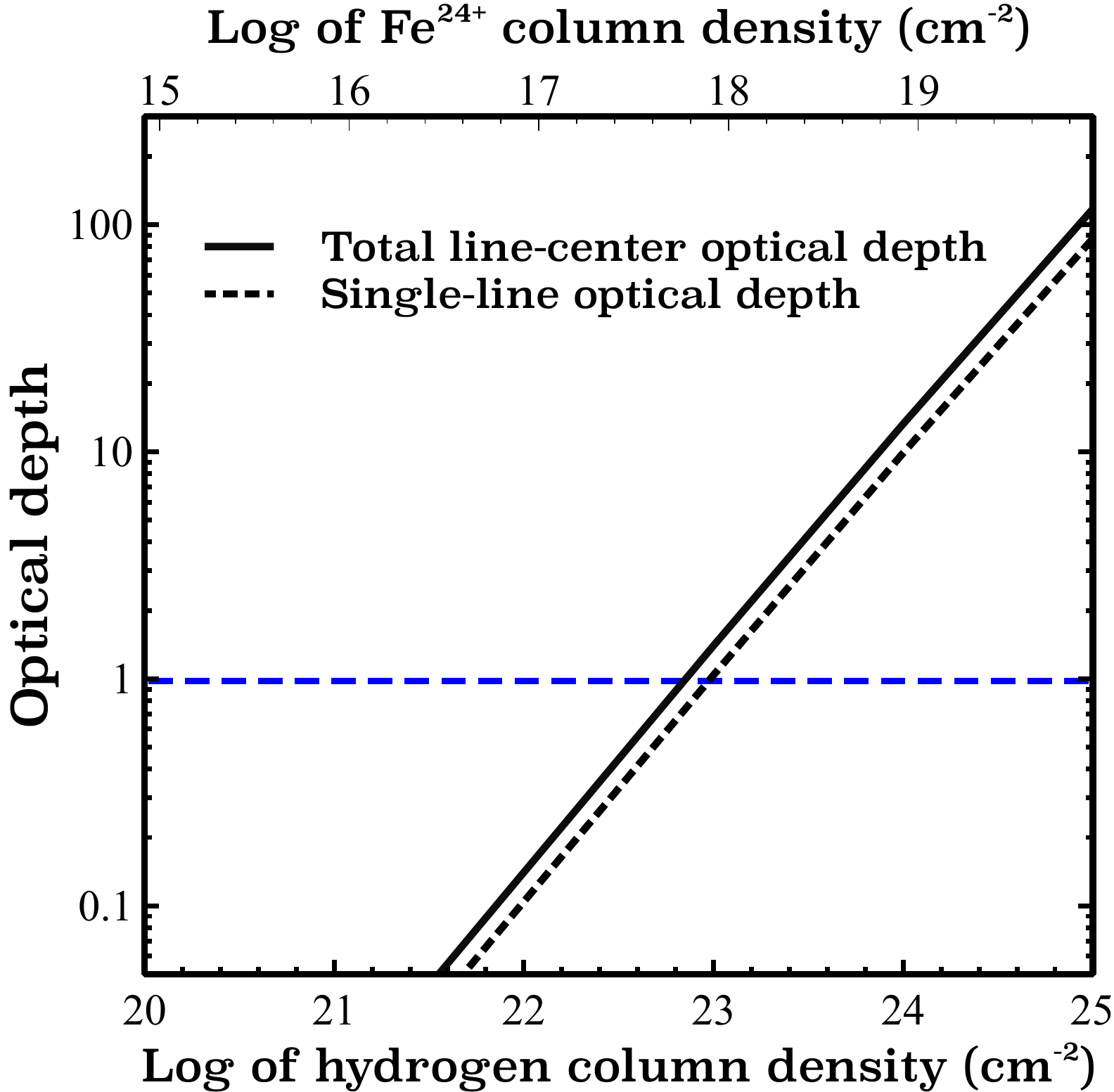}
\caption{
 Optical depth vs. hydrogen column density for $3\,\,^{3}P \rightarrow 1\,\,^{1}S$  transition in Fe XXV. Solid: Total line-center optical depth, Dotted: Individual line optical depth. 
The x-axis on top shows log of Fe$^{24+}$ column density. The blue dashed line marks the optical depth of unity.}
\label{fig:3to1opt}
\end{figure}
 
Unlike x, total line-center optical depth in y has a $\sim$ 60\% contribution from absorption by Fe$^{24+}$ itself. 
  When a y photon gets absorbed by Fe$^{24+}$, it leads to two possible modes of re-emission.
  It can either be re-emitted as a y photon or make a transition
  to $2\,\,^{3}S$ following the emission of a z photon. 
  The probability of re-emission via either of these modes depends on
  the transition probabilities   between the levels or, more specifically,
  branching ratios. The branching ratio for scattered y photon being re-emitted 
  as a $2\,\,^{3}P_{1}$ $\rightarrow$  $2\,\,^{3}S$ following a z 
  transition is quite small :
  \begin{equation*}
     \frac{A_{2\,\,^{3}P_{1} \rightarrow 2\,\,^{3}S}}  {{(A_{2\,\,^{3}P_{1} \rightarrow 2\,\,^{3}S}+ A_{2\,\,^{3}P_{1} \rightarrow 1\,\,^{1}S})}}  \sim 10^{-5}
 \end{equation*}
Therefore the majority of the scattered y photons will be re-emitted as y with no decrease in y intensity.
  This indicates a possible increase in z intensity, causing the slower drop of y/z ratio with increasing column density in Figure \ref{f:CaseAB2}.
  
 Similarly, $\sim$ 100\% of w photons are scattered by Fe$^{24+}$. Upon being scatted, there are two possible modes of 
  re-emission for w photos. The branching ratio for $2\,\,^{1}P$ $\rightarrow$  $2\,\,^{1}S$ following the transition to the ground is very small:
  
  \begin{equation*}
     \frac{A_{2\,\,^{1}P \rightarrow 2\,\,^{1}S}}  {{(A_{2\,\,^{1}P \rightarrow 2\,\,^{1}S}+ A_{2\,\,^{1}P \rightarrow 1\,\,^{1}S})}}  \sim 10^{-6}
 \end{equation*}
  Therefore almost all the scattered w photons will be re-emitted as w. The drop in the w/z ratio in Figure \ref{f:CaseAB2} mostly results from an increase in z intensity apart from the resonance scattering effects in w at large column densities.
 
  Even though none of the $n=2\rightarrow 1$ photon takes part in the Case A to B transfer, higher $n$ Lyman lines ($n= 3, 4, 5...\rightarrow 1$) show such transfer.
 The factor contributing to the increase in z is Case A to B transfer in $n= 3, 4, 5...\rightarrow 1$ photons. In an optically thick cloud, these photons are scattered, followed by emission of  Balmer line photons plus the  $n=2\rightarrow 1$ photons. This results in a surplus of $n=2\rightarrow 1$ photons. Tables 4.1 and 4.2 in \citet{2006agna.book.....O} show the Case A and Case B limit in HI recombination lines.  Here we discuss the increase in z intensity due to this process, whereas the slight increase in y intensity will be discussed in Section \ref{Measuring Column density}.

  Case A to B transfer for  $3^{3}P\rightarrow 1^{1}S$ transition is shown  in the upper panel of Figure \ref{f:3to1}. Lower panel shows Case A to B transition in $n= 4, 5, 6, 7, 8, 9, 10\rightarrow 1$ photons. A plot similar to that of 1b in paper I is shown in Figure \ref{fig:3to1opt} comparing the single line and total line-center optical depths for $3\,\,^{3}P \rightarrow 1\,\,^{1}S$  transition. Line ratios in Figure \ref{f:3to1} are taken relative to z for the reason explained previously.

  Once a photon for $3\,\,^{3}P \rightarrow 1\,\,^{1}S$ transition gets scattered,
  the probability that it will emit a $3\,\,^{3}P \rightarrow 2\,\,^{3}S$ following  
  the emission of a $2\,\,^{3}S \rightarrow 1\,\,^{1}S$ (z) photon is $\sim 32\%$ in one scattering:
  \begin{equation*}
     \frac{A_{3\,\,^{3}P \rightarrow 2\,\,^{3}S}}  {{(A_{3\,\,^{3}P \rightarrow 2\,\,^{3}S}+ A_{3\,\,^{3}P\rightarrow 1\,\,^{1}S})}}  \sim 0.32
 \end{equation*}

Therefore, for a single event of scattering, $3\,\,^{3}P \rightarrow 1\,\,^{1}S$ photons have a 32 \%  probability of being converted to z and Balmer series line photons. At a hydrogen column density of N$_{H}$= 10$^{25}$ cm$^{-2}$, the triplet to singlet transition photons will experience  multiple scatterings  because of their large optical depths ($\sim$ 100, see Figure \ref{fig:3to1opt}). The majority of these line photons will be converted to z and Balmer series photons at this column density. This contributes to the increase in the intensity of z line photons in Figure \ref{f:CaseAB2}, and a decrease in the intensity of $3\,\,^{3}P \rightarrow 1\,\,^{1}S$ line photons in the top panel of Figure \ref{f:3to1}. Likewise, $n= 4,5...10 \rightarrow 1$ transition line photons  show a similar behavior indicating a Case A to B transfer (bottom panel of Figure \ref{f:3to1}), and selectively contribute to the increase in z intensity. Although we show the Case A to B transfer of Lyman line photons up to n=10, we include collapsed levels up to n $\leq$ 100 in our simulation.
The net increase in the z line intensity will result from the collective Case A to B transfer of Lyman line photons for transitions from all the levels up to n=100.

\begin{figure}[h!]
\centering
\includegraphics[scale=0.5]{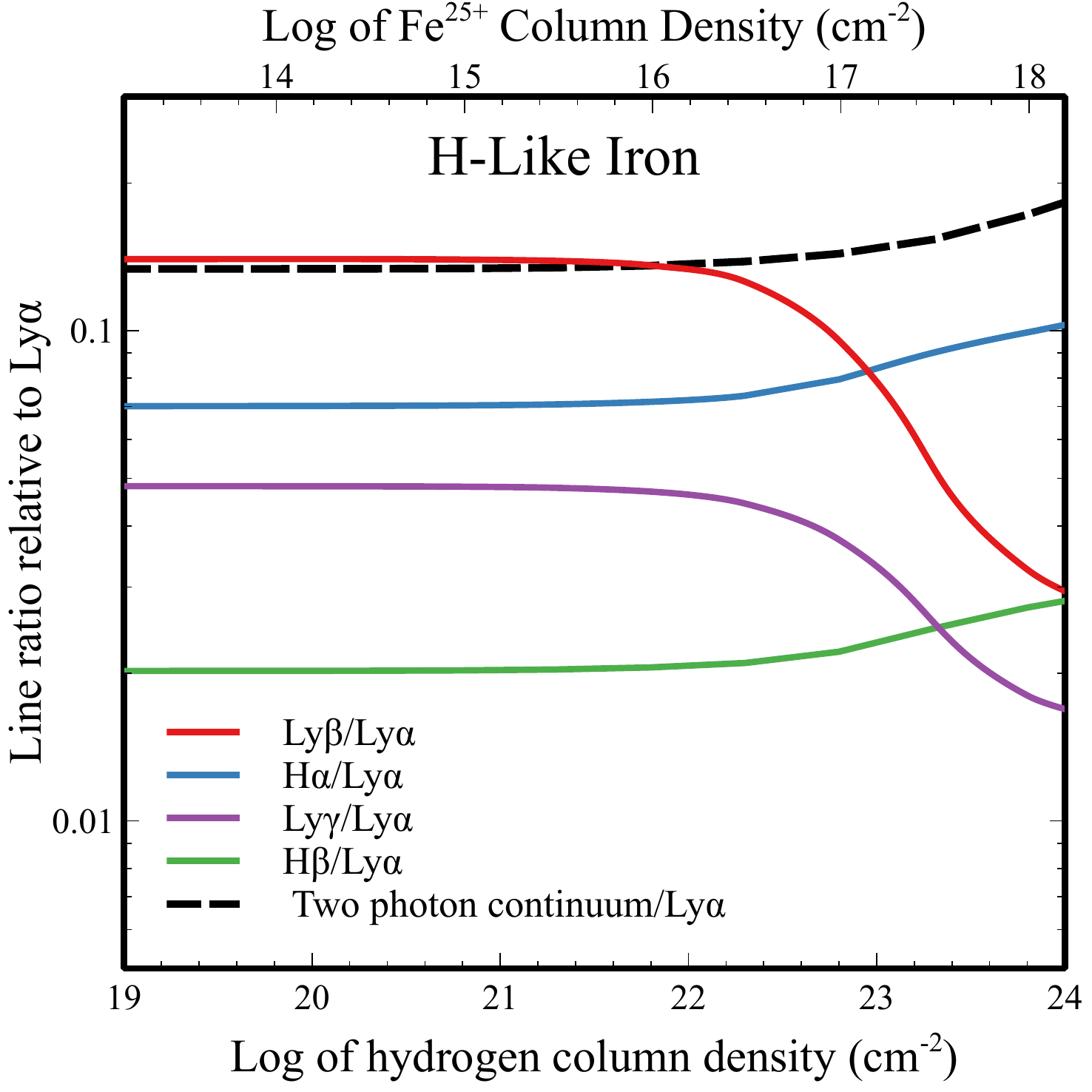}
\caption{Fe XXVI line intensity ratios with respect to Ly$\alpha$ vs. log of hydrogen column density.  The x-axis on top shows log of Fe$^{25+}$ column density. The dashed line is the integrated two-photon continuum relative to Ly$\alpha$.} 
\label{fig:FeXXVI_colden}
\end{figure}

\begin{figure*}
\gridline{\fig{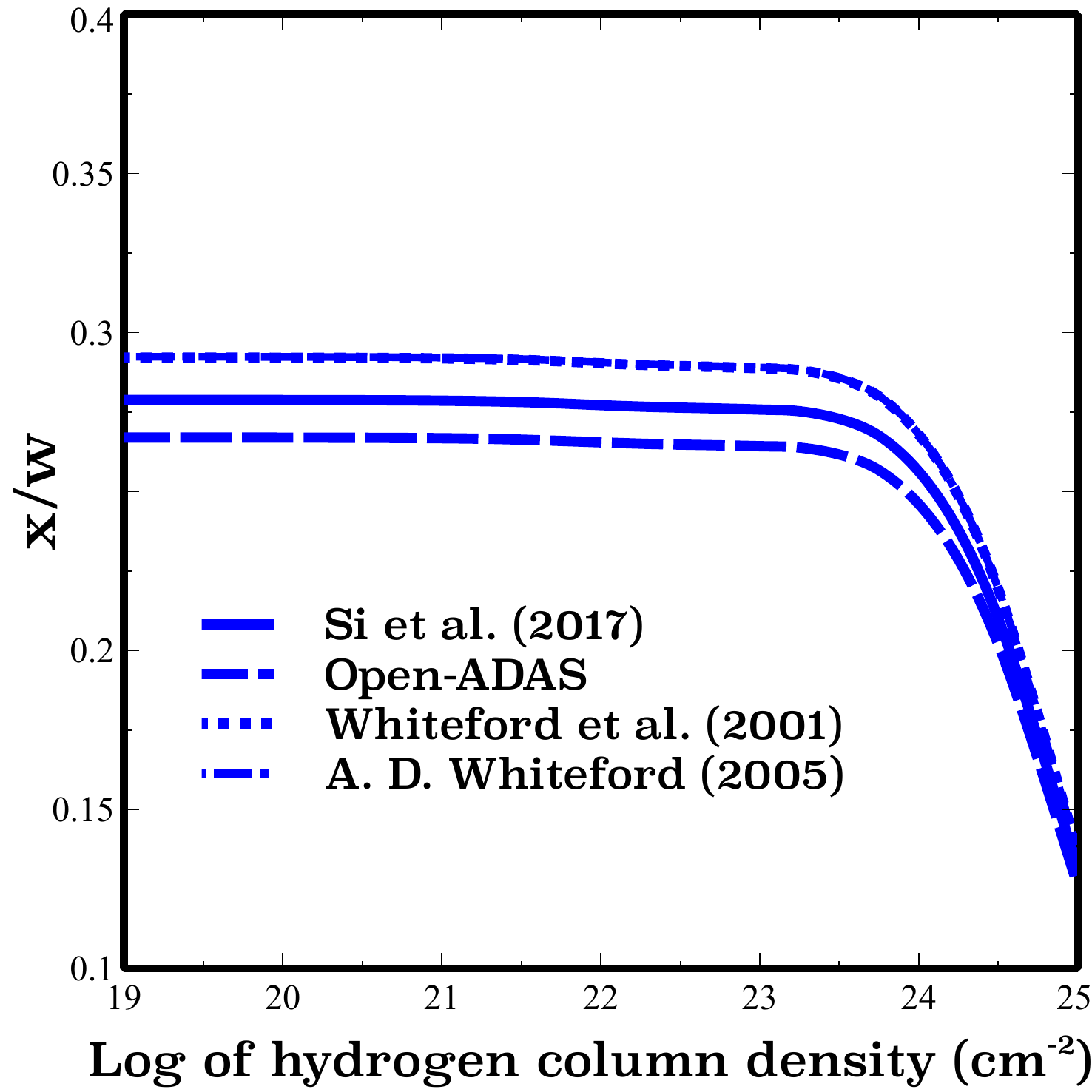}{0.32\textwidth}{(a)}
          \fig{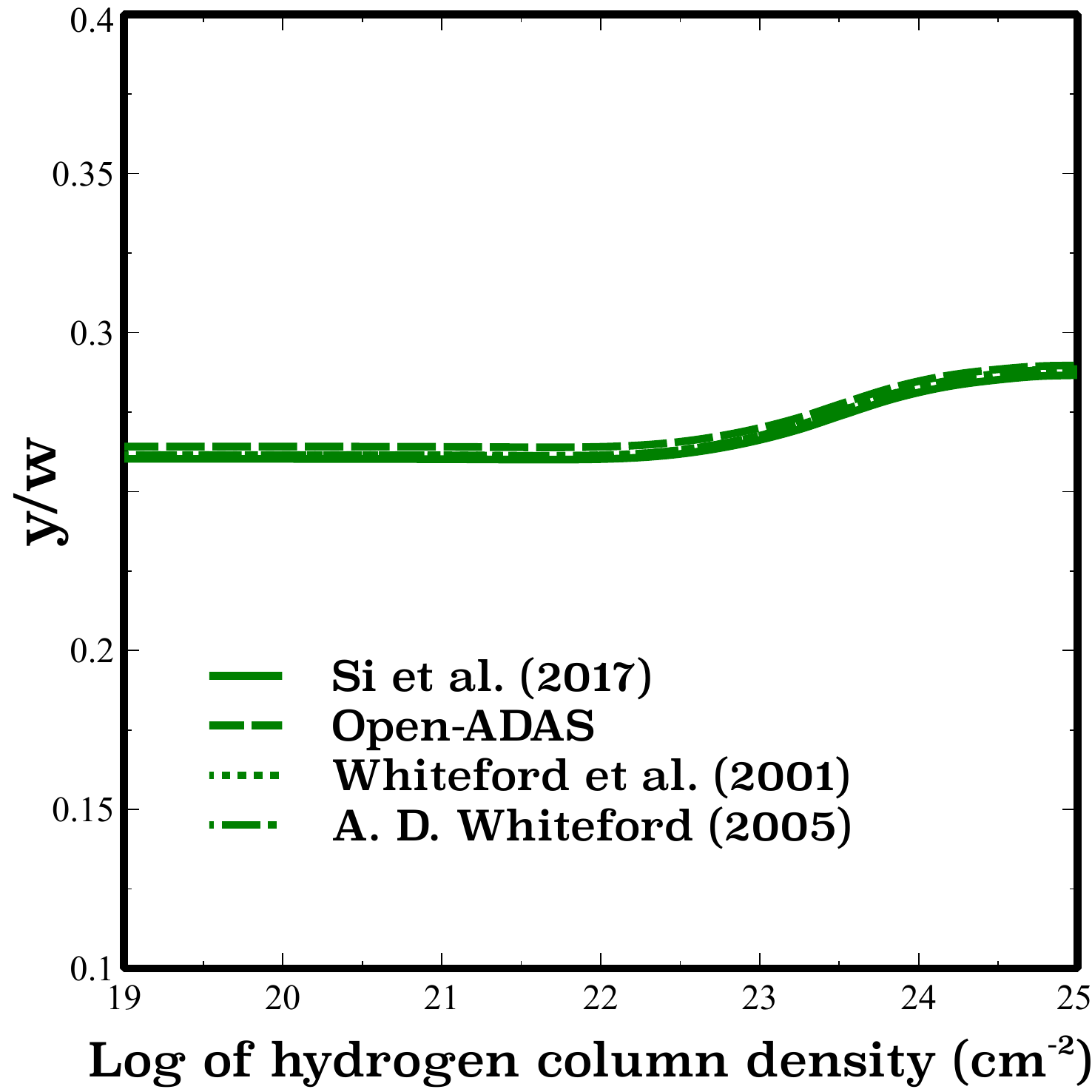}{0.32\textwidth}{(b)}
          \fig{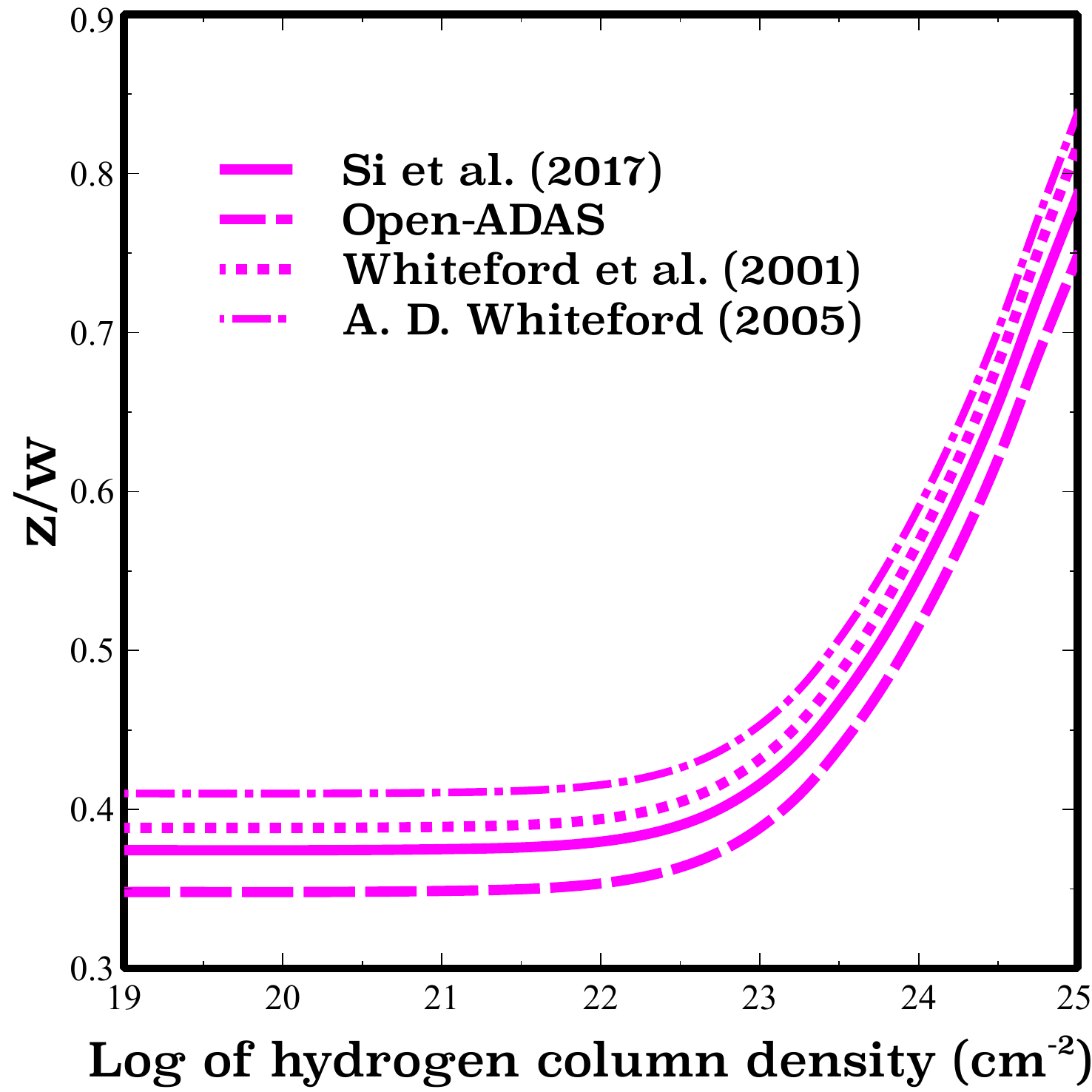}{0.32\textwidth}{(c)}
          }
\gridline{\fig{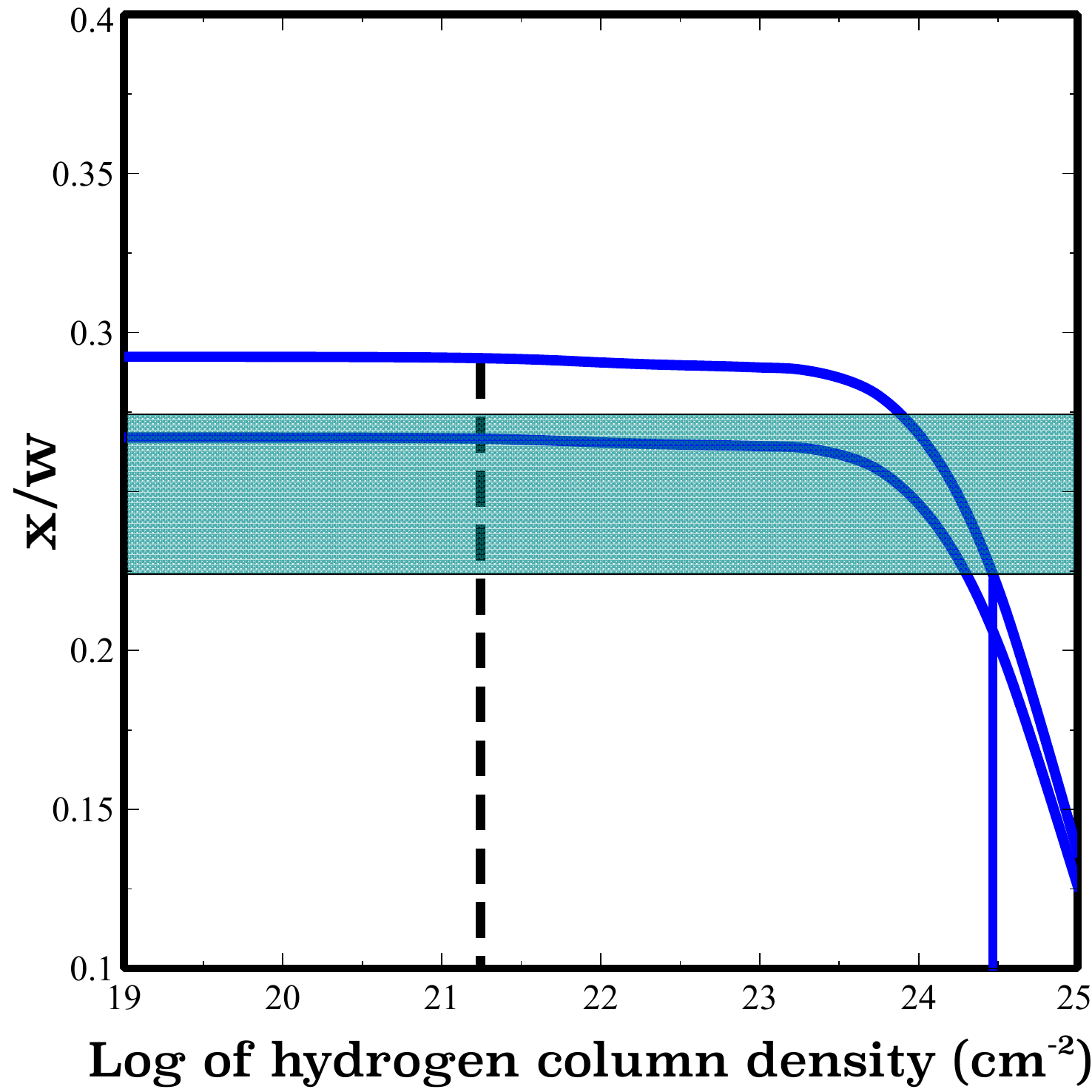}{0.32\textwidth}{(d)}
          \fig{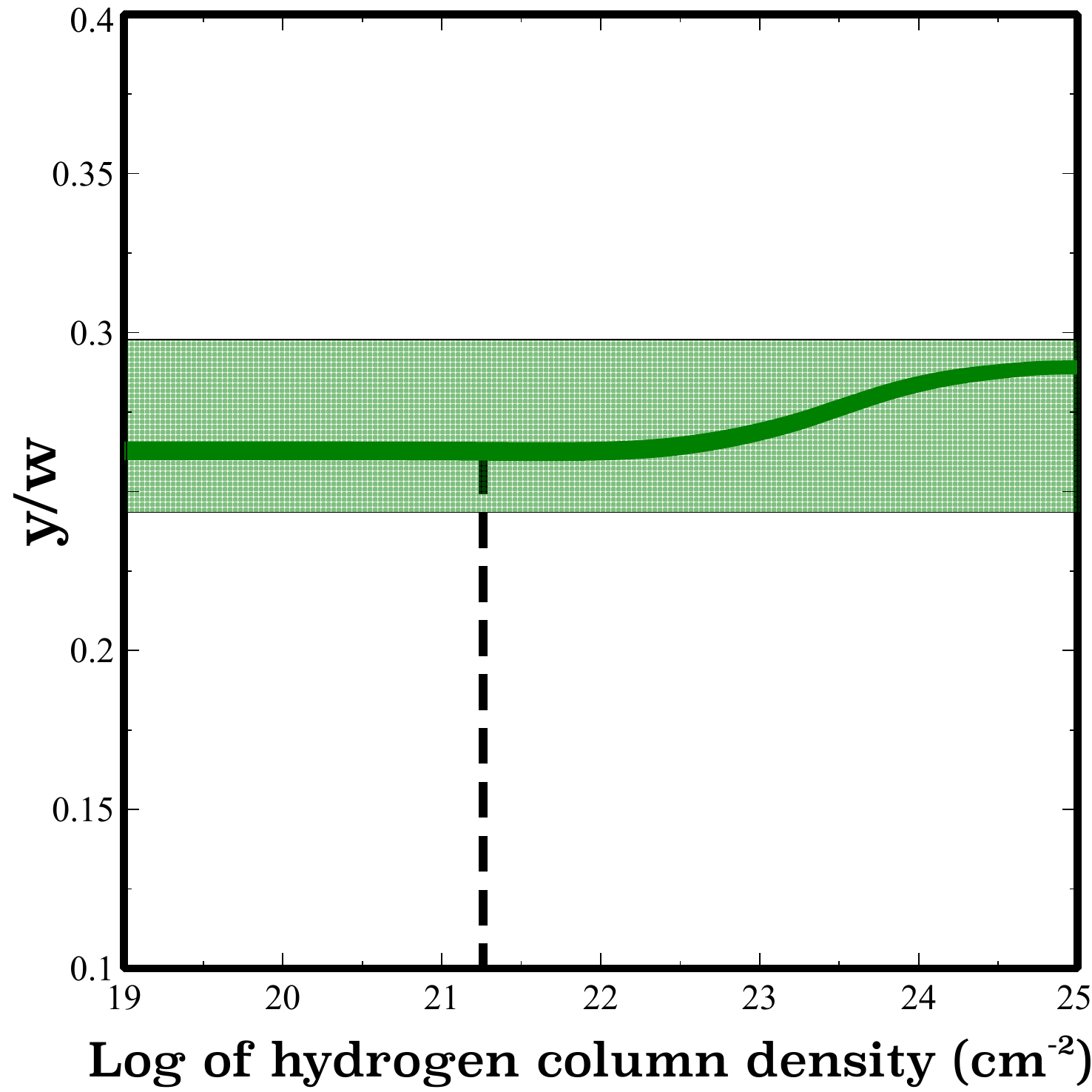}{0.32\textwidth}{(e)}
          \fig{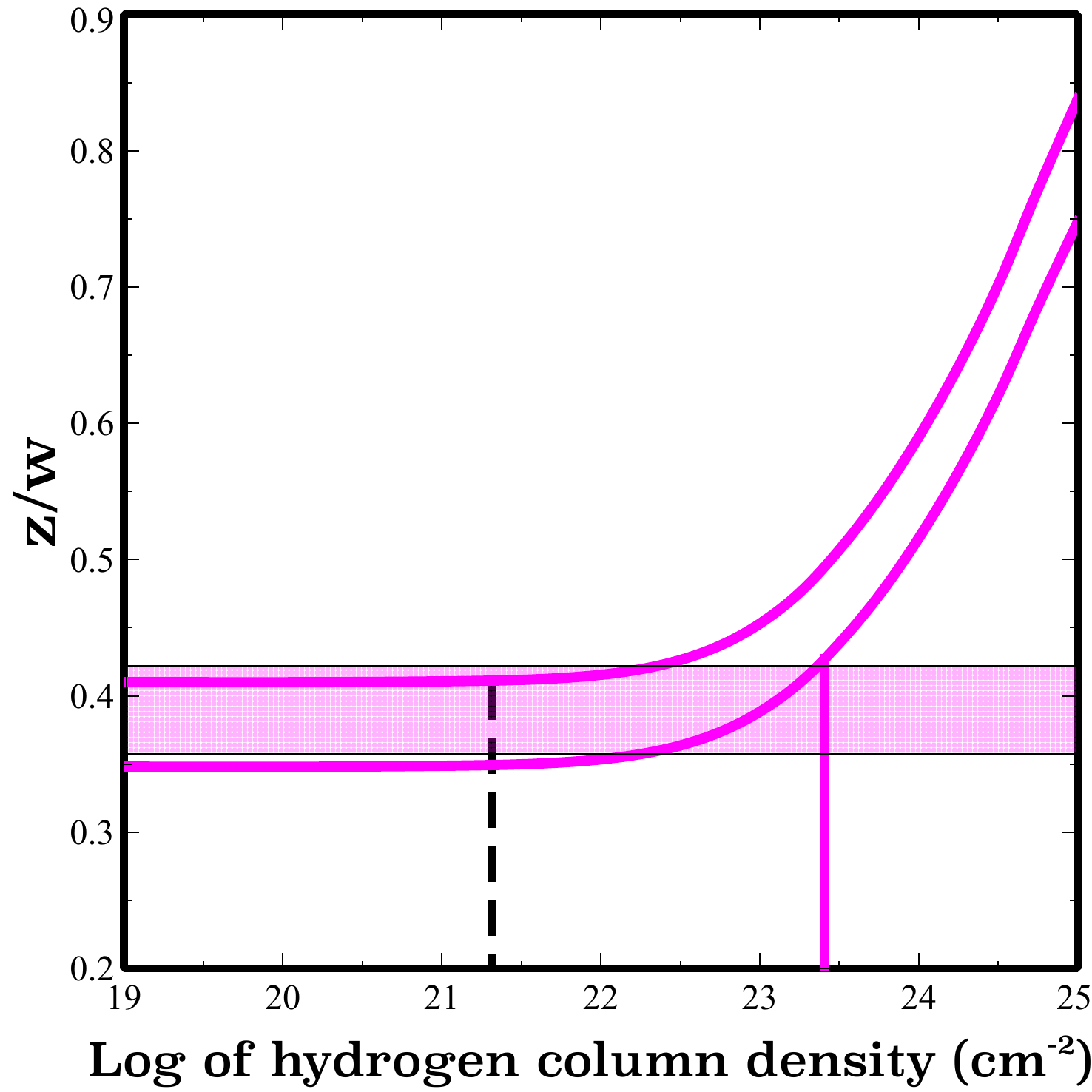}{0.32\textwidth}{(e)}
          }
\caption{Variation of x/w, y/w, and z/w line ratios with hydrogen column density. 
Top panel: Line  ratios calculated with CLOUDY for different collision datasets;  
Bottom panel: Line  ratios calculated with CLOUDY with their net uncertainties overplotted with the \textit{Hitomi} observed line ratios. Solid vertical lines show constraints on column density. \textit{Hitomi} reported hydrogen column density is shown with black dashed lines in each panel.
\label{f:col_den}}
\end{figure*}

\subsubsection{Fe XXVI}\label{Fe_XXVI}
As mentioned in Section \ref{X-ray emission from H-like Fe XXVI}, 
there is no fundamental difference between Fe$^{25+}$ and other hydrogenic ions. 
We show the Case A to B transition for the line intensity ratios
Ly$\beta$, Ly$\gamma$, H$\alpha$, and H$\beta$
with respect to Ly$\alpha$ in Figure \ref{fig:FeXXVI_colden}
(Ly$\alpha$, Ly$\beta$, Ly$\gamma$, H$\alpha$, and H$\beta$ are $n$=2$\rightarrow$1, $n$=3$\rightarrow$1, $n$=4$\rightarrow$1, $n$=3$\rightarrow$2, and $n$=4$\rightarrow$2 transitions with
wavelengths $\SI{1.78177}{\angstrom}$, $\SI{1.50337}{\angstrom}$, $\SI{1.42541}{\angstrom}$, $\SI{9.62154}{\angstrom}$, and $\SI{7.12706}{\angstrom}$, respectively).
The bottom and top axes in the Figure respectively show the hydrogen and Fe$^{25+}$ column densities.
Line ratios are plotted instead of the individual line intensities for the same reason, as discussed in Section \ref{Fe XXV}.

Figure \ref{fig:FeXXVI_colden} shows that line ratios remain unchanged at small column densities
when the lines are optically thin (Case A). 
But at higher column densities, the cloud becomes optically thick, the 
spectrum goes to Case B,
and the higher $n$ Lyman lines like Ly$\beta$ and Ly$\gamma$
degrade to  H$\alpha$, and  H$\beta$ respectively. This causes the Lyman lines
 to become weaker, and  the Balmer lines to become stronger. Such analysis can be extended to other hydrogenic ions like Si XIV,
 S~XVI, Ar XVIII, and Ca XX for constraining/measuring column density. This will be explored in future papers. 

\subsection{Constraints on Column density}\label{Measuring Column density}
We run a line ratio diagnostic for Fe XXV K$\alpha$ complex with column density
using the Case A to B transition.  Here we plot the variation of Fe XXV line 
ratios (RS corrected) relative to the strongest resonance (w) line with increasing hydrogen column density 
 in Figure \ref{f:col_den}. 
 
At lower column densities,  lines ratios remain constant with 
the increase in column density. Apart from the RS effects at the higher column densities,
line intensities in w increase linearly, while z and x increase 
faster and slower than linear, respectively. 
The behavior of x line intensity with column density is discussed in paper I.  Faster than linear growth in z is due to the Case A to B transfer in selective $n=3,4...\rightarrow 1$  photons (See section \ref{Fe XXV}).

Similar to the increase in z/w, the slight increase in y/w in the high column density can also be explained with Case A to B transfer in $n=3,4...\rightarrow 1$ line photons to generate Balmer line photons plus the  $n=2\rightarrow 1$ transition photons. Such transition makes all the $n=2\rightarrow 1$ transitions brighter in the Case B limit compared to Case A. This process strengthens the weak $n=2\rightarrow 1$ transitions more than  w, the strongest, leading to the slight rise in y/w at the high-column-density limit.

The top panel of Figure \ref{f:col_den} shows the variation of line ratios relative to w with hydrogen column density 
for different collision datasets. 
The bottom panel of the Figure compares the variation in line ratios calculated with CLOUDY, with the observed line ratios
 to get constraint on column density.  
 The shaded horizontal regions are the observed line ratios with their uncertainties at the reported column density by \citet{2018PASJ...70...12H}.
 Section \ref{Line ratios from Hitomi observations} describes
 the extraction of line ratios along with their errors from the observed spectra. 
 The regions between the two solid lines in all three panels
 enclose the calculated line ratios with CLOUDY. 
 The enclosed regions are inclusive of the uncertainty in temperature
 and collision strengths coming from different collision datasets.
 Refer to Table \ref{t:cs_Fe XXV} for the uncertainties in collision datasets.

 The intersection regions between observed and calculated line ratios 
 for x/w and z/w predict a hydrogen column 
 density of $ N_{\rm H} \leq  2.95\times 10^{24} \pscm$, and $ N_{\rm H} \leq  2.75\times 10^{23} \pscm$, respectively. The solid vertical lines in the bottom panel of Figure \ref{f:col_den} show the upper limit of column density from these two ratios. The y/w ratio calculated with CLOUDY intersect with the observed value for all column densities below $10^{25} \pscm$. These upper limits calculated from x/w, y/w, and z/w ratios are
 all consistent with the reported hydrogen column density $ N_{\rm H,hot} \sim  1.88\times 10^{21} \pscm$ (shown with the black vertical dashed lines)
 for Perseus.

\subsection{Effects of turbulence}\label{Effects of turbulence}

\begin{figure}[h!]
\centering
\includegraphics[scale=0.5]{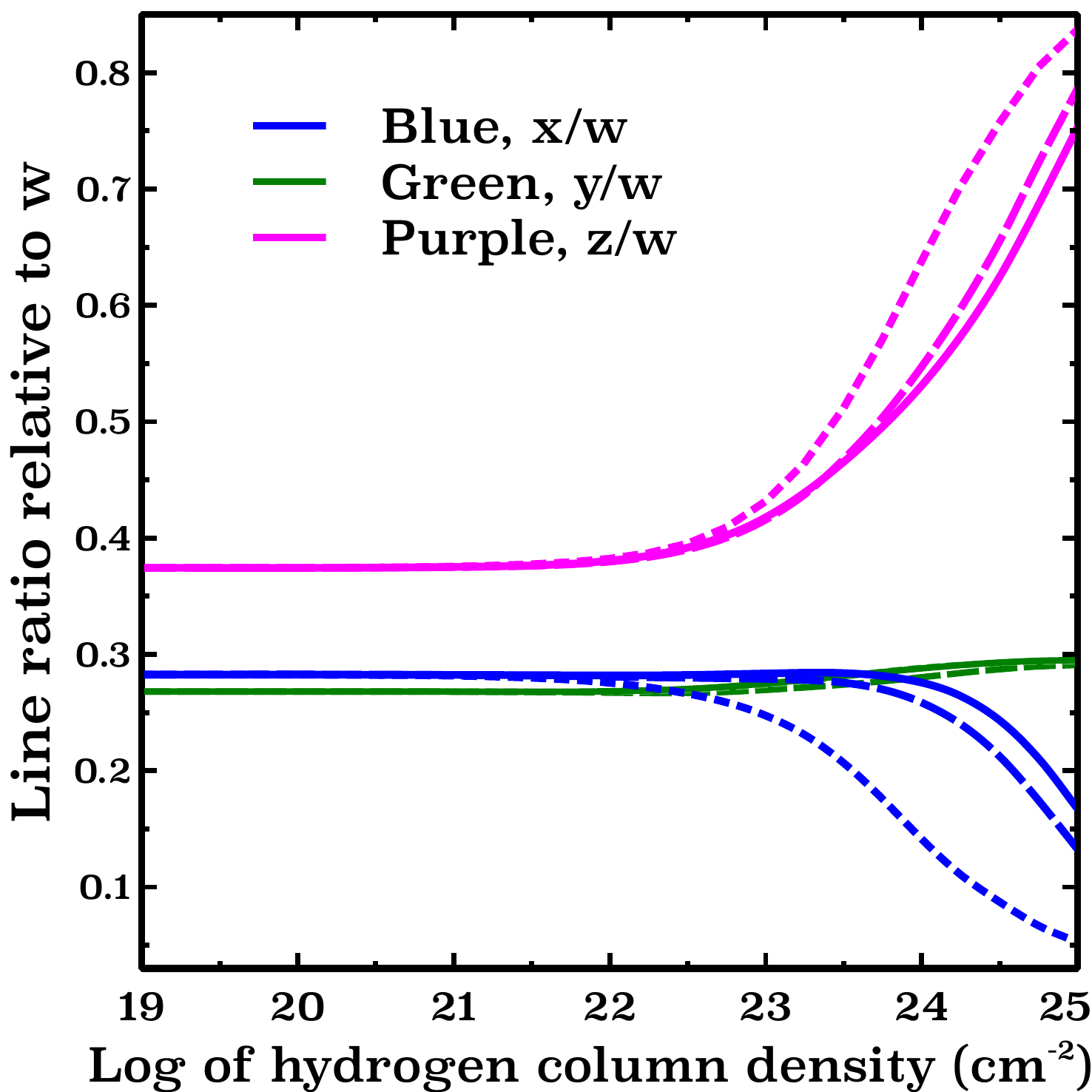}
\caption{Line intensity ratio relative to w vs. log of hydrogen column density. Solid: no turbulence, Dashed: Turbulence 150 km/s, Dotted: Turbulence 500 km/s.}
\label{fig:turbulence}
\end{figure}

In Figure \ref{fig:turbulence}, we show the effect of turbulence on the Fe XXV K$\alpha$ complex line ratios.
\citet{2018PASJ...70...10H} found a difference in turbulent broadening between 
w (159–167 km/s), and x, y, z (136–150 km/s) in Obs23:out.
We use a turbulent velocity of 150 km/s for simplicity in our simulation 
of the outer region of Perseus core. 
Figure \ref{fig:turbulence} shows that
such a quiescent gas environment barely affects the line ratios,
as the difference in the line ratios calculated at the turbulence 
0 km/s and 150 km/s is very small.

\begin{figure*}
\gridline{\fig{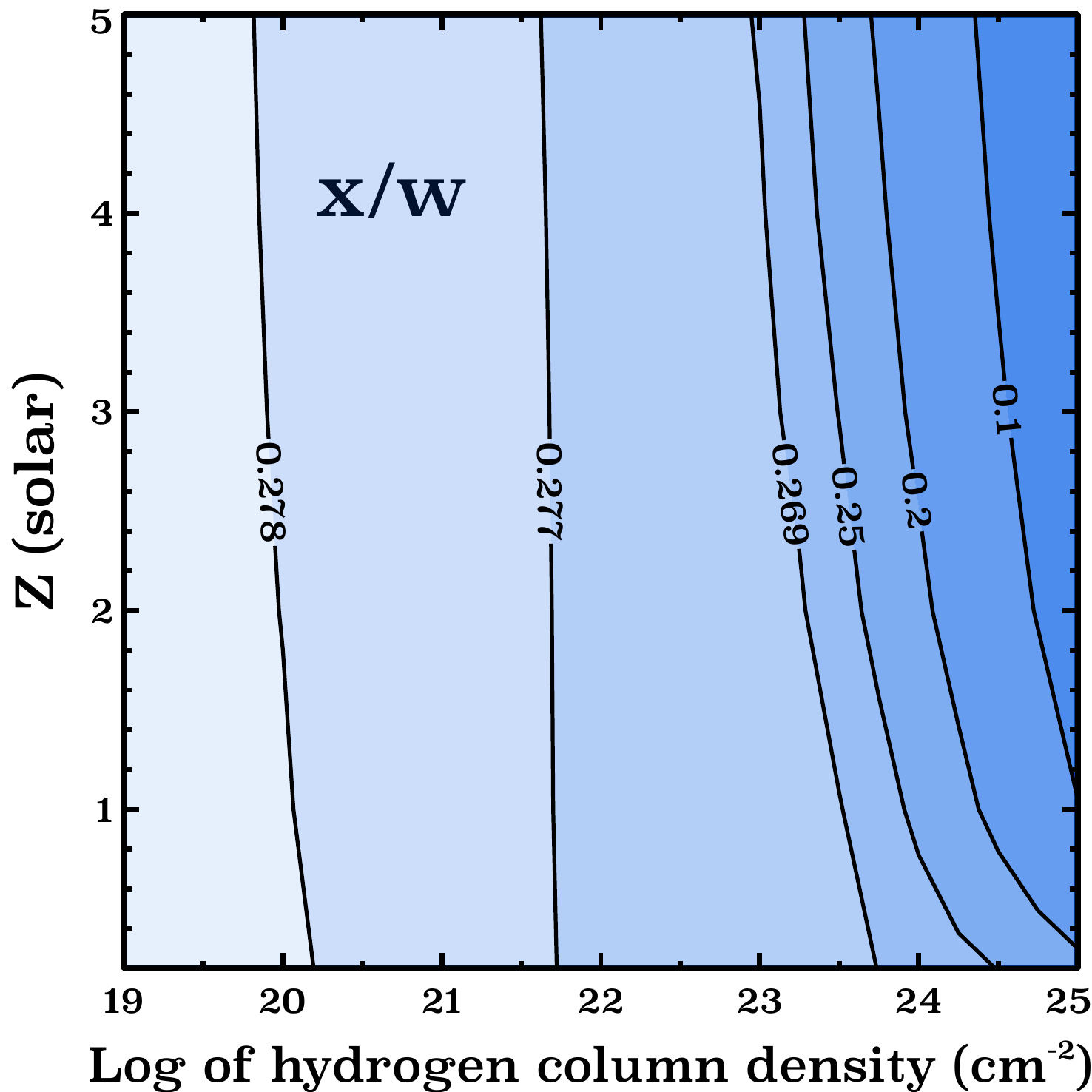}{0.32\textwidth}{(a)}
          \fig{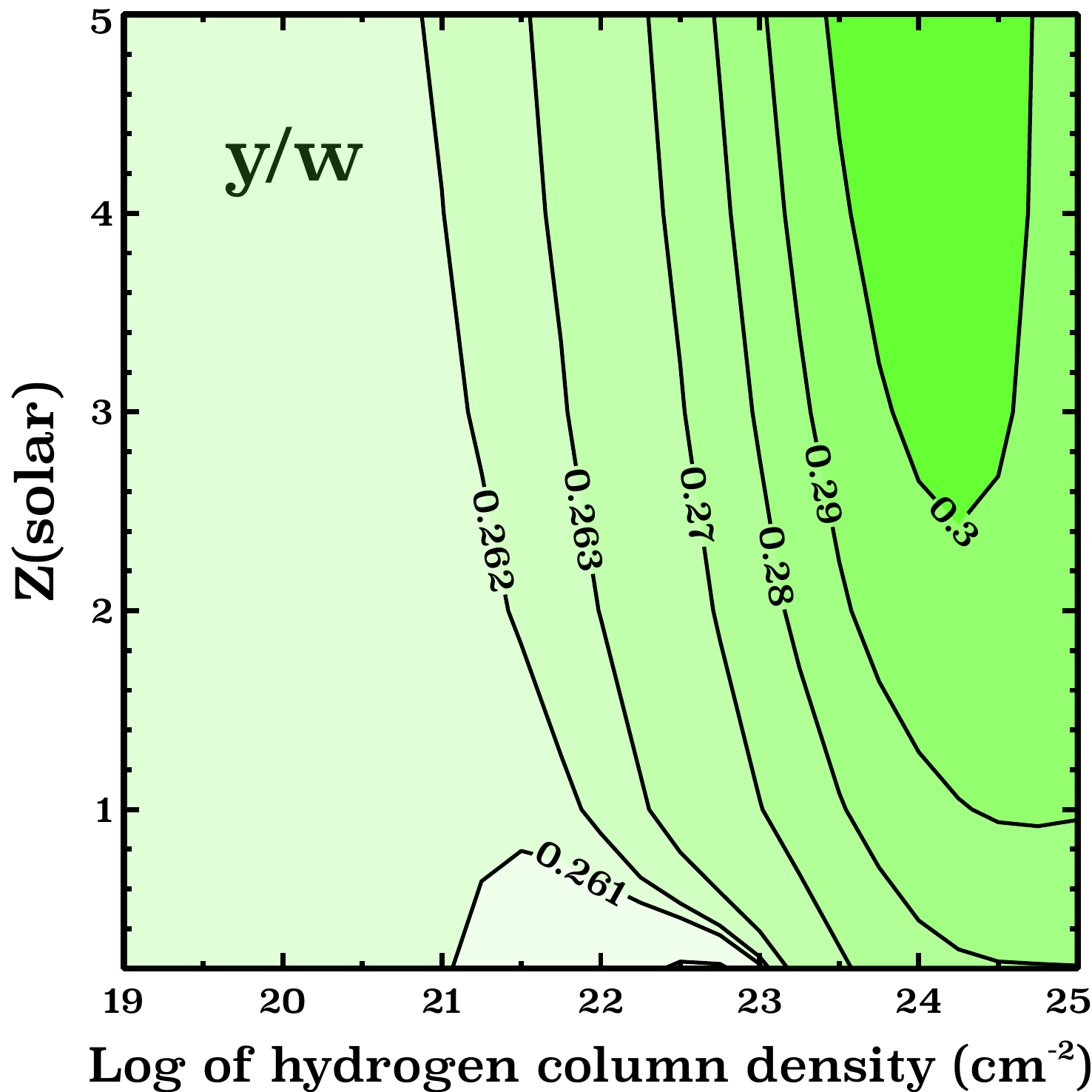}{0.32\textwidth}{(b)}
          \fig{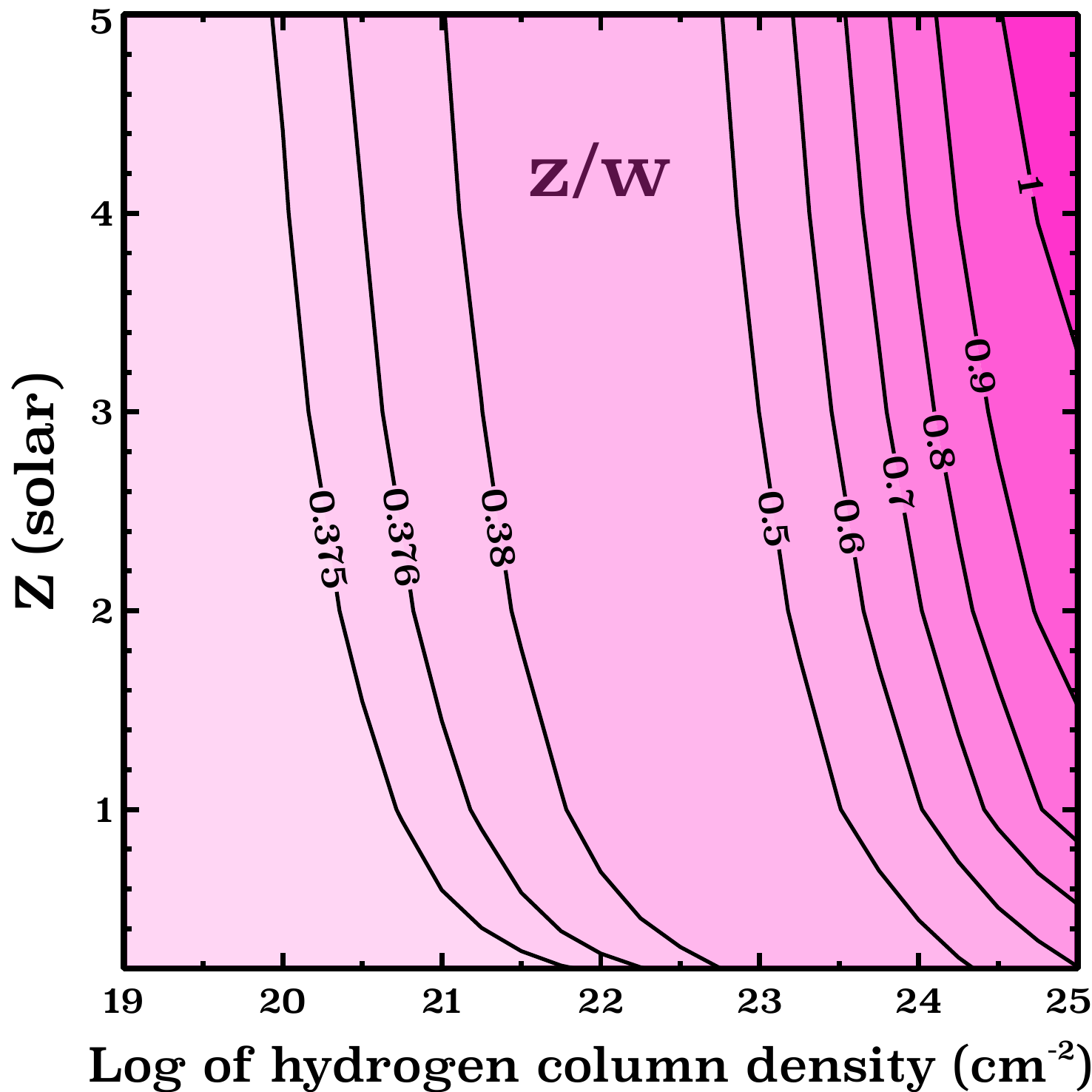}{0.32\textwidth}{(c)}
          }

\caption{Left, middle and right panels  show x/w, y/w, and z/w as a function of hydrogen column density and metallicity, respectively.  
\label{f:contourplot}}
\end{figure*}

\begin{figure*}
\gridline{\fig{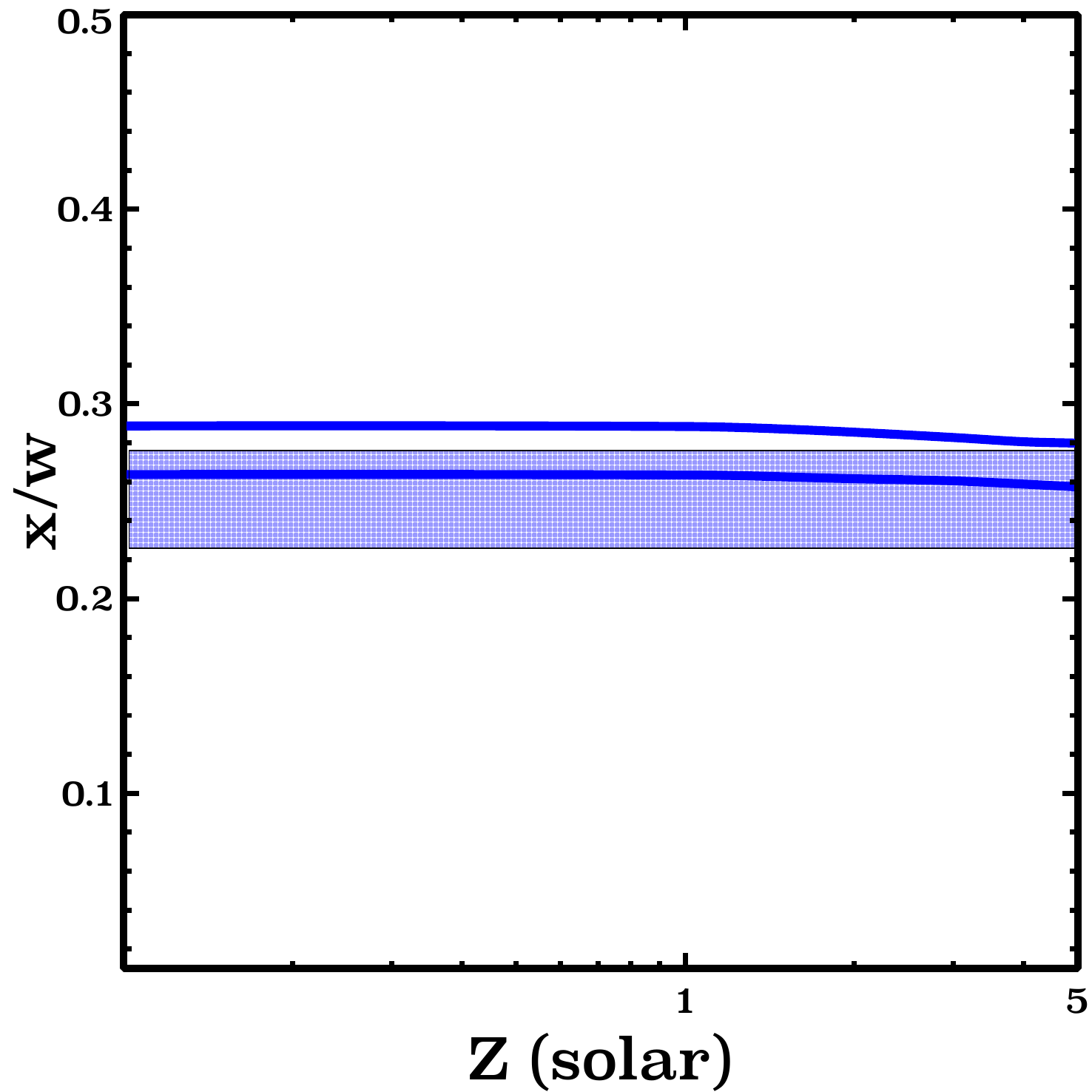}{0.32\textwidth}{(a)}
          \fig{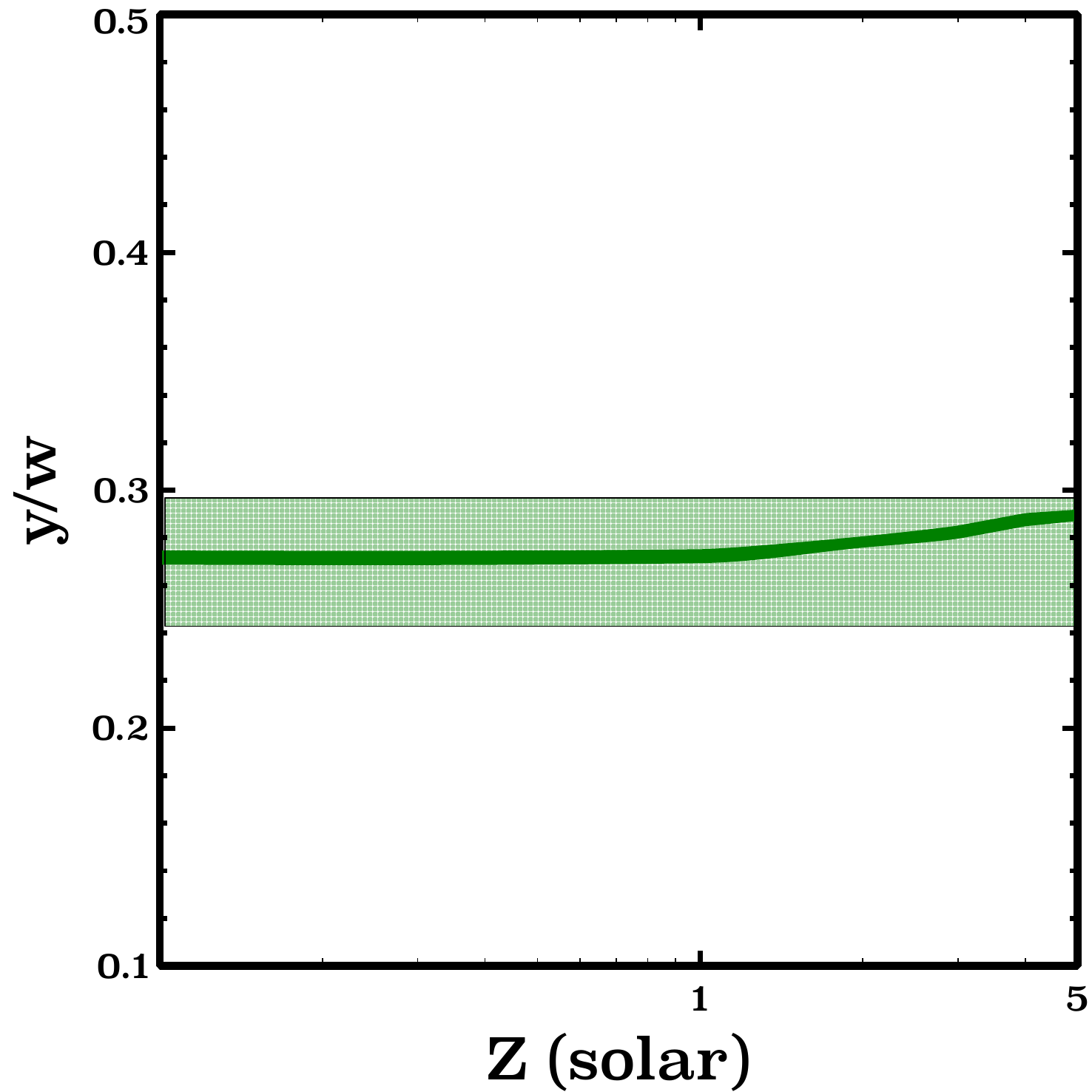}{0.32\textwidth}{(b)} 
          \fig{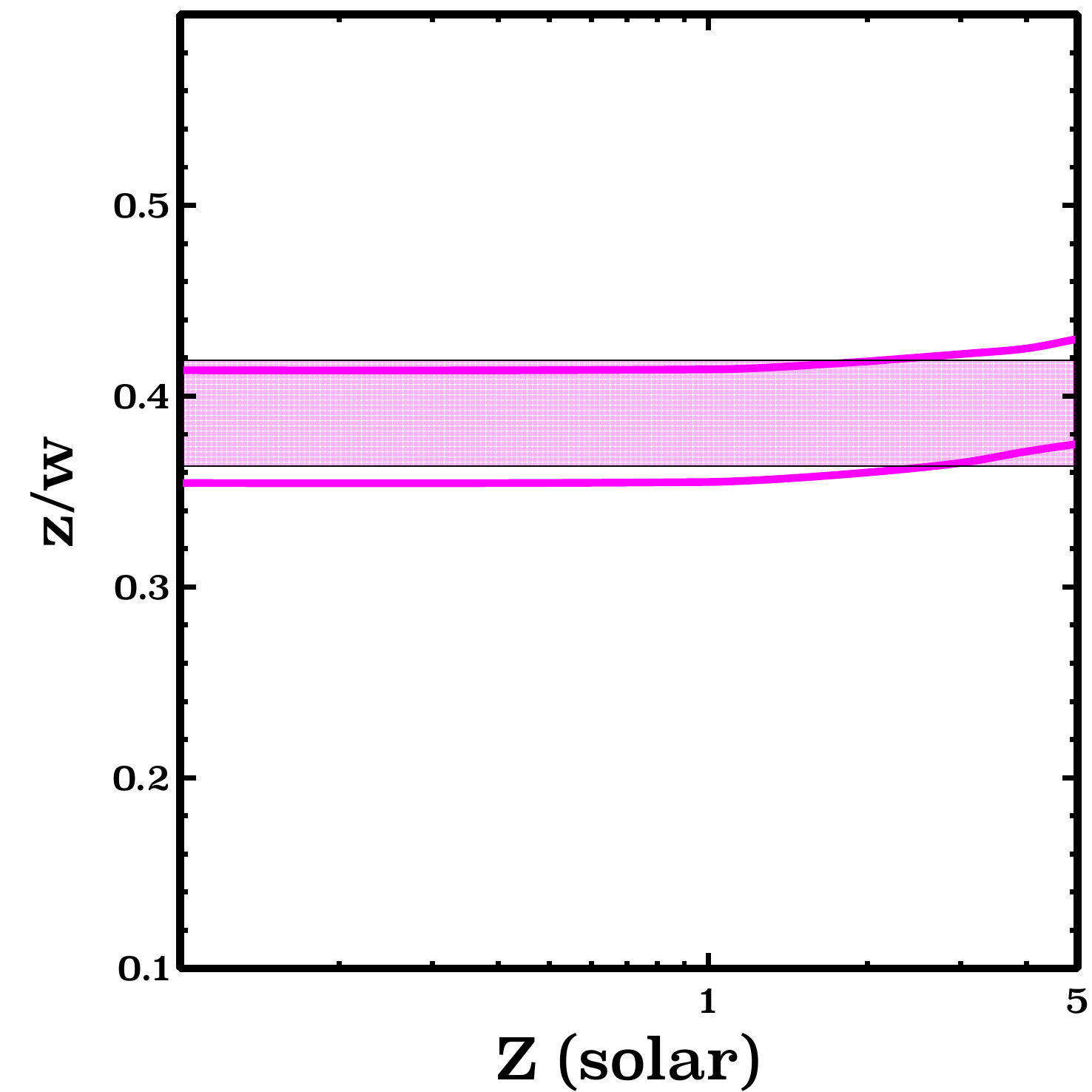}{0.32\textwidth}{(c)}
          }

\caption{Line intensity ratios versus metallicity at $ N_{\rm H} =  1.88\times 10^{21} \pscm$. Blue, green, purple lines in left, middle, and right panels enclose the x/w, y/w, and z/w calculated with CLOUDY. The enclosed regions are inclusive of the uncertainty in temperature, and different collision datasets. Observed line ratios with errors are shown with the horizontal bars in all three panels.   
\label{fig:Metallicity}}
\end{figure*}
We also plot the line ratios at the turbulence of 500 km/s. 
Such a high value of the turbulence is not applicable in our region of interest,
but may apply for gas in other environments such as merging galaxy clusters \citep{2005MNRAS.357.1313C}.

Optical depth can be expressed as a product of column density ($N$),
and absorption cross-section ($\alpha_\nu$):
\begin{equation}\label{e:2}
\tau_\nu = N \alpha_\nu
\end{equation}
\noindent
where $\alpha_\nu$ is inversely proportional to the total Doppler velocity.
\begin{equation}\label{eqn:LineAbsorptionCrossSection}
\begin{split}
{\alpha _\nu(x) }  & = \frac{{{\lambda ^{3}}g_u}}{8 \pi g_{l}} \frac{A_{u,l}}{{\pi^{1/2} {u_{\rm Dop}}}}{\varphi _\nu }(x) \\ 
 & = \frac{{{\pi ^{1/2}}q_e^2\lambda
\,{f_{l,u}}}}{{{m_e}c{u_{\rm Dop}}}}{\varphi _\nu }(x)\big[\rm cm^{2}\big]
\end{split}
\end{equation}

where $A_{u,l}$ is the downward transition probability, $f_{l,u}$ is the oscillator strength and other symbols have their usual meaning \citep{1970stat.book.....M}. 
If the temperature is held constant,
an increase in the turbulent velocity decreases $\alpha_\nu$,
which decreases the overall optical depth. 

Thus ideally, to achieve the same optical depth at zero and higher turbulent velocity, 
a higher column density should be required for the latter case. However, this outcome gets reversed  due to line interlocking of x, y, and z with other ions (e.g., Fe$^{23+}$, Cr$^{22+}$, and Fe$^{22+}$; see paper I). The higher the turbulent velocity is, the higher the Doppler width of a line and the more pronounced the line interlocking effects will be. Total line-center optical depth will, therefore, increase with turbulent velocity rather than decrease, as equation \ref{eqn:LineAbsorptionCrossSection} suggests.

\begin{figure}[h!]
\centering
\includegraphics[scale=0.5]{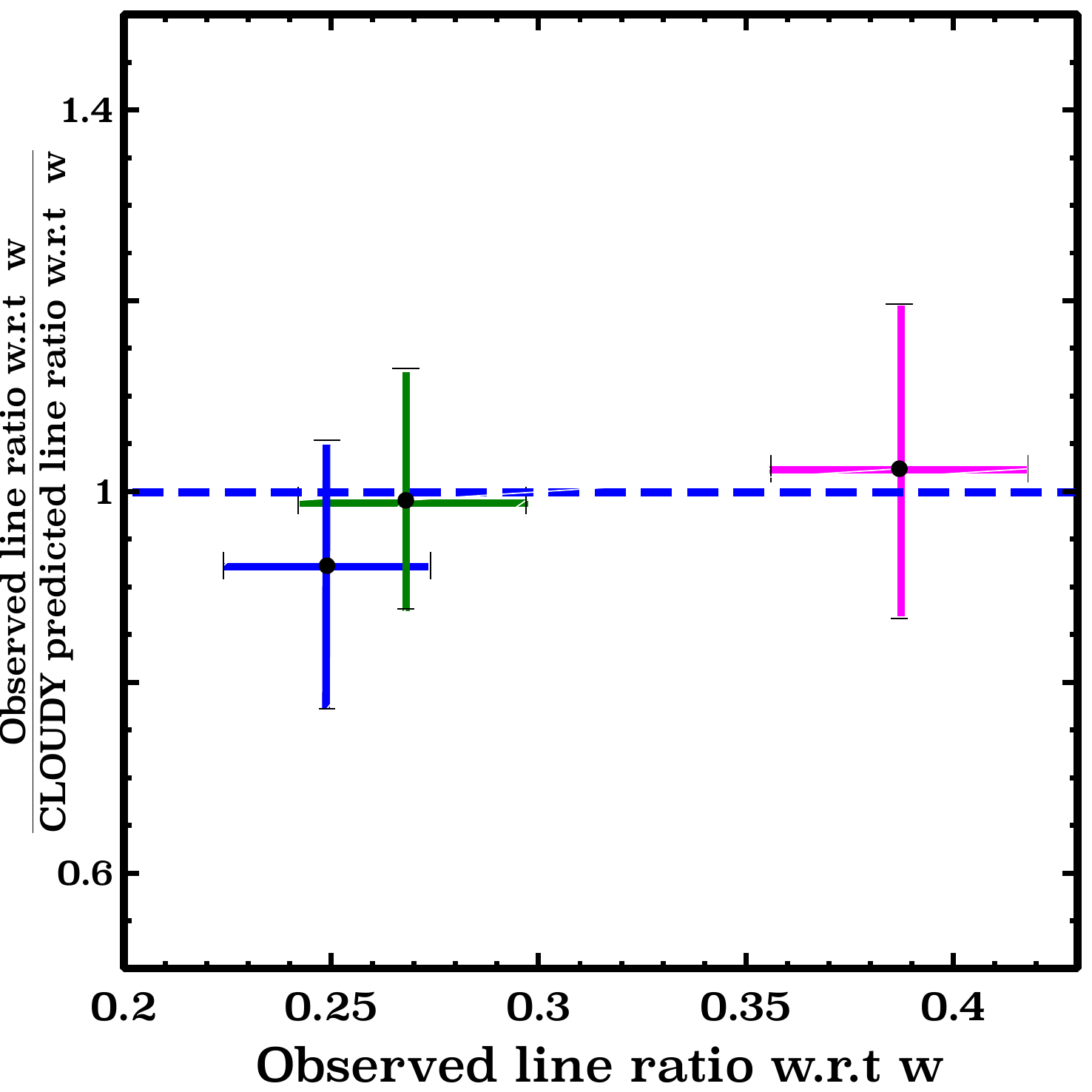}
\caption{The ratio of observed to CLOUDY predicted line ratios plotted as a function of observed line ratios.
All the ratios are taken with respect to the resonance (w) line. 
The points with red, blue, and purple error bars show the degree of agreement
with the \textit{Hitomi} observed line ratios for x/w, y/w, and z/w, respectively. Vertical error bars are calculated using 
observational and CLOUDY estimated uncertainties.
The blue dotted line shows the ideal situation where observed 
line ratios exactly match with the predicted. }
\label{fig:discussion}
\end{figure}

 \begin{table*}
\centering{
  \caption{\label{t:cs_Fe XXV}List of collision strengths at 4 keV from different atomic datasets for ${n= 2\rightarrow 1}$ transitions in Fe XXV.}}
  
\begin{tabular}{llrrrrr}
\hline

\multicolumn{5}{r}{Collision Strengths} \\
\cline{3 -7}
Label & Transition    & Whiteford et al. (2001) & A. D. Whiteford (2005) & A. Giunta (2012) & Si  et  al.(2017) & Difference \\
\hline
 w&$2^{1}P\rightarrow1^{1}S$ &3.97e-3 & 3.99e-3 &4.32e-3 &  4.18e-3  &  8\%   \\
 x&$2^{3}P_{2}\rightarrow1^{1}S$ &7.27e-4  &7.39e-4  &6.72e-4  & 7.13e-4 & 9\%         \\
 y&$2^{3}P_{1}\rightarrow1^{1}S$ &7.24e-4 & 7.31e-4 & 8.68e-4 & 7.90e-4 &   17\%     \\
 -&$2^{3}P_{0}\rightarrow1^{1}S$ &1.48e-4 & 1.48e-4 & 1.35e-4 & 1.51e-4 &  11\%      \\
 z&$2^{3}S\rightarrow1^{1}S$ &3.05e-4 &4.28e-4 &2.46e-4 & 3.16e-4 &    42\%    \\
 -&$2^{1}S\rightarrow1^{1}S$ &8.41e-4 &8.59e-4 &8.68e-4 & 1.05e-3 &   20\%    \\
\hline
\end{tabular}
 \end{table*}
\subsection{Interplay between column density and metallicity}\label{Interplay between column density and metallicity}

Figure \ref{f:contourplot} shows  the line ratios relative to w as contour plots by 
varying the hydrogen column density and metallicity.
In the low-column-density limit,  the line ratios are minimally affected by change in metallicity.  
Figure \ref{fig:Metallicity} shows the effects of metallicity on the line ratios 
at the reported value  of hydrogen column density by \citet{2018PASJ...70...12H} ($ N_{\rm H,hot} \sim  1.88\times 10^{21} \pscm$). The uncertainties in temperature and  collision strengths from the sources mentioned previously contribute to the uncertainties in 
 the line ratios (shown with the solid enclosed lines in Figure \ref{fig:Metallicity}). At the reported hydrogen column density, x, y, and z are optically thin (paper I). Therefore, the line ratios are minimally affected with change in metallicity as these lines are still optical thin for the metallicity range  (0$\leq$ Z(solar)$\leq$5) shown in the Figure.
 
\citet{2018PASJ...70...10H} 
 use deprojected heavy element abundances relative to solar in the inner 
 150 kpc region from Chandra data archive for a spherically symmetric Perseus model.
 At a distance of $30 - 60$ kpc from the central AGN of Perseus core, 
 the reported range in metallicity is  $\sim0.5-0.75$ of solar.
(see Figure 7 in their paper). 
We overplot the CLOUDY calculated line ratios with \textit{Hitomi} observed line ratios, and obtain a region of overlap for all the metallicities between 0 and 5.  This is consistent with the reported metallicity range for the outer region of Perseus core, but calculating/constraining the best-fit metallicity requires an extension of parameter space up to higher column densities.

In the high-column-density limit, the line ratios exhibit a clear variation with metallicity (see Figure \ref{f:contourplot}). It can be seen that a smaller metallicity is equivalent to a larger hydrogen column density for the same optical depth. In systems with hydrogen column densities greater than $10^{23} \pscm$, like Seyfert 2 Galaxies \citep{1999ApJ...522..157R, 2001ApJ...560..139T, 2011ApJ...729...30M}, such comparison between the calculated and observed line ratios can be used to measure/constrain metallicities.

\section{Summary}\label{Discussion}
New generations of micro-calorimeter on board \textit{Hitomi} 
made precision spectroscopy in X-ray possible for the first time.
This enables us to explore radiative transfer effects like 
Case A to B transition in X-ray, previously observed in optical / UV \citep{1938ApJ....88...52B}.
In this paper, we document the Case A to B transition for H-like and He-like
iron and show that this can be used to constrain column density.
We also show the effect of and turbulence, and interplay between column density and metallicity in the outer region
of the Perseus core using a line ratio diagnostic. In addition,
we show that highly charged He-like systems behave differently from He I 
and other low charge He-like systems. This is due to the atomic number dependence of transition probabilities,
leading to fast transitions in non-allowed triplet to 
singlet transitions. 

From a comparison between the CLOUDY predicted and \textit{Hitomi} observed line ratios in the outer region of Perseus core for the observed temperature, column density, and metallicity (see Figure \ref{fig:discussion}),
we find an agreement of $\sim 91\%$, $99\%$, $98\%$
in the x/w, y/w, and z/w ratios. 

In the case B limit, two factors contribute to the variation in line ratios with column density for the same sets of physical parameters. First, line photons get absorbed in the cloud and are re-emitted as different lines (see Section \ref{Fe XXV}). Second, line photons are absorbed in the cloud and re-emitted in a different direction leading to the change in intensity of line photons along our line of sight (i.e., resonance scattering). Physically, it represents the migration of photons from the center of the cluster, where there is a photon deficit, to the outer regions, where there would be a photon surplus \citep{1987SvAL...13....3G}.  Our current model treats the cluster as a single sphere and can not recover this physics region wise. A proper model of the variation in density and temperature across the cluster will be investigated in a later paper. The first line to become optically thick is w (see Figure 1b in paper I). We use a flux suppression factor of $\sim$ 1.28 for w reported by \citet{2018PASJ...70...10H} at the outer region of Perseus core for the best-fit column density $ N_{\rm H,hot} \sim  1.88\times 10^{21} \pscm$ \citep{2018PASJ...70...12H}.  Flux suppression in x, y, and z is insignificant at such small column density but needs to be considered for higher column densities when these lines become optically thick. 

In Section \ref{Measuring Column density}, we introduced a novel method of measuring/constraining column density from Case A to B transition. In the case of Perseus, x, y, and z are still in the optically thin regime (Case A). Therefore, we only get an upper limit in column density instead of a specific value. However, this method will be useful in systems with higher column densities ($ N_{\rm H} \geq  10^{23} \pscm$) in the optically thick regime (Case B) for measuring column density.





\section*{Acknowledgment}

    The comments of the referee were very helpful and added significantly to the presentation of our work.  His or her help is gratefully acknowledged. We thank Anna Ogorzalek and Stefano Bianchi for their valuable comments.
We acknowledge support by NSF (1816537, 1910687), NASA (17-ATP17-0141, 19-ATP19-0188), and STScI (HST-AR-15018).
MC also acknowledges support from STScI (HST-AR-14556.001-A).

\software{Xspec (v12.10.1, \citet{1996ASPC..101...17A}), HEAsoft (v6.25, \citet{2014ascl.soft08004N}), CLOUDY \citep{2017RMxAA..53..385F}}


\bibliography{references}{}
\bibliographystyle{aasjournal}

\end{document}